\makeatletter
\@ifundefined{@parse@version@dash}{%
\def\@parse@version#1{\@parse@version@0#1}
\def\@parse@version@#1/#2/#3#4#5\@nil{%
\@parse@version@dash#1-#2-#3#4\@nil}
\def\@parse@version@dash#1-#2-#3#4#5\@nil{%
  \if\relax#2\relax\else#1\fi#2#3#4 }
}{}
\makeatother
\documentclass[reprint, amsmath,amssymb, prl]{revtex4-2}

\usepackage{tikz}
\usepackage{color}
\usepackage{ulem}
\usepackage{graphicx}
\usepackage{dcolumn}
\usepackage{bm}
\usepackage[colorlinks, linkcolor=blue, anchorcolor=blue, citecolor=blue, urlcolor=blue]{hyperref}
\usepackage{url}
\definecolor{lime}{HTML}{A6CE39}
\DeclareRobustCommand{\orcidicon}{%
	\begin{tikzpicture}
	\draw[lime, fill=lime] (0,0)
	circle [radius=0.16]
	node[white] {{\fontfamily{qag}\selectfont \tiny ID}};
	\draw[white, fill=white] (-0.0625,0.095)
	circle [radius=0.007];
	\end{tikzpicture}
	\hspace{-2mm}
}

\foreach \x in {A, ..., Z}{%
	\expandafter\xdef\csname orcid\x\endcsname{\noexpand\href{https://orcid.org/\csname orcidauthor\x\endcsname}{\noexpand\orcidicon}}
}

\begin{document}

\title{Spatial Non-Locality Induced Non-Markovian EIT in a Single Giant Atom}
\author{Y. T. Zhu\orcidA{}$^{1,2,3}$}
\author{R. B. Wu$^{4,5}$}
\author{S. Xue\orcidB{}$^{1,2,3}$}
\email{shbxue@sjtu.edu.cn}
\affiliation{$^{1}$Department of Automation, Shanghai Jiao Tong University, Shanghai 200240, P. R. China}
\affiliation{$^{2}$Key Laboratory of System Control and Information Processing, Ministry of Education of China, Shanghai 200240, P. R. China}
\affiliation{$^{3}$Shanghai Engineering Research Center of Intelligent Control and Management, Shanghai 200240, P. R. China}
\affiliation{$^{4}$Department of Automation, Tsinghua University, Beijing 100084, P. R. China}
\affiliation{$^{5}$Beijing National Research Center for Information Science and Technology, Beijing 100084, P. R. China}
\date{\today}

\begin{abstract}
In recent experiments, electromagnetically induced transparency (EIT) were observed with  giant atoms, but nothing unconventional were found from the transmission spectra. In this letter, we show that unconventional EIT does exist in giant atoms, and indicate why it has not been observed so far. Different from these existing works, this letter presents a consistent theory including a real space method and a time delayed master equation for observing unconventional EIT. We discover that this phenomenon is a quantum effect which cannot be correctly described in a semi-classical way as those in recent works. Our theory shows that it can be observed when the time delay between two neighboring coupling points is comparable
to the relaxation time of the atom, which is crucial for a future experimental observation. This new phenomenon results from inherent non-locality of the giant atom, which physically forces propagating fields to be standing waves in space and the atom exhibiting retardations in time. Our theory establishes a framework for application of nonlocal systems to quantum information processing.
\end{abstract}

\maketitle

\noindent{\textbf{Introduction---}} In recent years, remarkable progresses in nano-acoustic resonators enable to investigate the photon-phonon quantum coherence at single-quanta level \cite{Gustafsson2014, Schuetz2015, Chu2017, Chu2018, Satzinger2018, Ekstrom2019, Manenti2017, Andersson2019}, where artificial atoms can be coupled piezoelectrically to acoustic waves  \cite{Manenti2017, Andersson2019}, or electrically  to microwaves  \cite{Kannan2020} at several distant points. Subsequently, this kind of artificial atoms is called giant atoms  \cite{Kockum2020} since their inherent non-locality resulted from multi-point couplings dominates their dynamics and unconventional phenomena  different from those for  small atoms are observed. For instance, non-locality makes the excitation spectra of a single giant atom exhibit unconventional atomic-frequency-dependent Lamb shifts and decay rates  \cite{Kockum2014, Gustafsson2014}, and also induces a non-exponential decay with non-Markovianity \cite{Guo2017, Andersson2019}. Besides, atomic excitations can counterintuitively oscillate without decay  \cite{Guo2020}. Also, several giant atoms with proper arrangements can be decoherece-free when they  dissipatively interact with a shared waveguide \cite{Kockum2018, Kannan2020}.

These works preliminarily discover some special features of two-level giant atoms, however, a full and exact description of giant atoms induced by non-locality is absent so that it would hinder further experimental discoveries. For example, only standard electromagnetically induced transparency (EIT) \cite{Harris1990, Boller1991, Fleischhauer2005} was  observed in a ladder-type giant atom with either acoustic couplings \cite{Andersson2020} or microwave couplings \cite{Vadiraj2021}. Especially, in Ref. \cite{Vadiraj2021}, without considering the non-locality as a quantum effect, the experiment data has an obvious discrepancy from the fit calculated by a semi-classical master equation. Hence, whether it is possible to discover EIT with special features for a three-level giant atom is still an open problem.

In this letter, we present a consistent theory with the real-space scattering method and the time-delayed master equation, with which we discover unconventional EIT phenomena. With our theory, non-locality is fully characterized in space and time,  which are represented by multi-peak scattering spectra and retarding effects, respectively. Also, we find the undesired spontaneous emission can be eliminated rather than suppressed by engineering the multiple-point-coupling structure such that the $\Lambda$-type configuration can be obtained from a $\Delta$-type artificial atom. In addition, we show that the propagating waves inside the outermost coupling points behave as standing waves due to interference between bidirectional propagating modes.

\noindent{\textbf{{Model---}} We consider a $\Delta$-type giant atom with levels $|1\rangle$, $|2\rangle$, and $|3\rangle$, and the corresponding transition frequencies $\omega_{21}$, $\omega_{31}$ and $\omega_{32}$. Here, the energy of the ground state $|1\rangle$ is assumed to be zero as a reference. The transitions $|3\rangle\leftrightarrow|1\rangle$ and $|2\rangle\leftrightarrow|1\rangle$ are respectively side coupled to waveguides $A$ and $B$ at multiple points $x_n$ and $\tilde x_m$ with indices $n=1,2,...,N, m=1,2,..,M$, and distances $d, \tilde{d}$, as shown in Fig. \ref{model}. The coupling strengths at these points are denoted as $g_{31}(x_n)=g_{31}e^{i\alpha x_n}$, $g_{21}(\tilde{x}_m)=g_{21}e^{i\beta \tilde{x}_m}$ with wavevectors $\alpha$ and $\beta$ for modes in the waveguide $A$ and $B$, respectively. Here, we take the waveguide $A$ as a channel of a probe field for observing its transmission and the waveguide $B$ as a reservoir leading to the spontaneous emission in the transition $|2\rangle\leftrightarrow|1\rangle$. Here, we consider the modes in both waveguides are photons, but the following results can be applied to phonons when the atom couples to surface acoustic waves via multiple interdigital transducers. To induce EIT phenomenon, a classical driving field with an angular frequency $\nu_c$ and a Rabi frequency $\Omega_c$ is applied to the transition $|2\rangle\leftrightarrow|3\rangle$.

\begin{figure}
\centering
\includegraphics[width=0.45\textwidth]{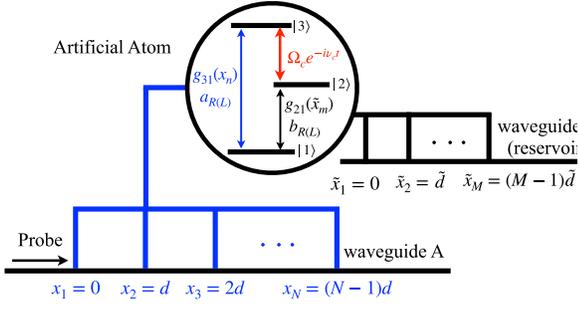}
\caption{Schematic of a $\Delta$-type giant atom. The transition $|1\rangle\leftrightarrow|3(2)\rangle$ is coupled to the waveguide $A$ ($B$) via multiple coupling points with an equal distance $d(\tilde{d})$. $g_{31}(x_n)$ and $g_{21}(\tilde{x}_m)$ denote the position-dependent coupling strength. Here, the waveguide $B$ is taken as a reservoir resulting in the undesired spontaneous emission of the atom, while we take the waveguide $A$ as a channel of a probe field. A classical driving with an angular frequency $\nu_c$ and a Rabi frequency $\Omega_c$ is applied to the transition $|2\rangle\leftrightarrow|3\rangle$ to induce EIT.}
\label{model}
\end{figure}

Different from the recent works~\cite{Andersson2020,Vadiraj2021}, the probe field herein is quantized. Under rotating wave approximation at the frequency $\nu_c$, an effective Hamiltonian of the system in real space reads
\begin{equation}
H_{\mathrm{eff}}=H_a+H_w+H_i,\label{Heff}
\end{equation}
where $H_a=(\omega_{21}+\nu_c-\frac{i\gamma_2}{2})|2\rangle\langle2|+(\omega_{31}-\frac{i\gamma_3}{2})|3\rangle\langle3|+\left(\Omega_c |3\rangle\langle2|+H.c.\right)$,  $H_w=\sum_{l=L, R}\int\mathrm{d}\tilde{x} b_l^\dag(\tilde{x})(\nu_c-if_l\tilde{v}_g\frac{\partial}{\partial \tilde{x}})b_l(\tilde{x})-\sum_{l=L, R}f_l\int\mathrm{d}x a_l^\dag(x)(iv_g\frac{\partial}{\partial x})a_l(x)$, and $H_i=\tilde{g}_{31}\sum_{l=L, R}\sum_{n=1}^N\int\mathrm{d}x \delta(x-x_n){\Big(}|3\rangle\langle1|a_l(x)+H.c.{\Big)}+\tilde{g}_{21}\sum_{l=L, R}\sum_{m=1}^M\int\mathrm{d}\tilde{x} \delta(\tilde{x}-\tilde{x}_m){\Big(}|2\rangle\langle1|b_l(\tilde{x})+H.c.{\Big)}$ are the Hamiltonians for the atom, the waveguides, and their interactions. We set $\hbar=1$, $f_R=1, f_L=-1$, and the group velocity of the field in the waveguide $A$ ($B$) is $v_g ~(\tilde{v}_g)$. The bosonic annihilation operators $a$ and $b$ with subscripts $R$ and $L$ are for the right- and left-going modes, respectively. The damping rates $\gamma_{2}$ and $\gamma_{3}$  denote non-waveguide losses of the atom \cite{Shen2009}. The Dirac $\delta$ function $\delta{\big(}x(\tilde{x})-x_n(\tilde{x}_m){\big)}$ indicates interacting positions. For simplicity, we have assumed that the coupling strengths at each coupling point are identical and denoted as $\tilde{g}_{3(2)1}$, and the linear dispersion relation holds in both waveguides. See Refs. \cite{SM2020, Shen2009} for more details of derivation.

\noindent{\textbf{Single-photon scattering---}} To investigate EIT at a single photon level, we assume that both the waveguides and the atom are initially prepared in their ground states $|\mathrm{vac}\rangle$ and $|1\rangle$. A single photon is incident from $x_1$ to $x_N$ and thus its scattering eigenstate $|\Psi\rangle$ in single-excitation subspace \cite{Jia2017, Bradford2012} is written as
\begin{eqnarray}
|\Psi\rangle =&&\int \mathrm{d}x \left[\phi_R^\alpha(x)a_R^\dag(x)+\phi_L^\alpha(x)a_L^\dag(x)\right]|\mathrm{vac}, 1\rangle+\nonumber\\
&&\int \mathrm{d}\tilde{x} \left[\phi_R^\beta(\tilde{x})b_R^\dag(\tilde{x})+\phi_L^\beta(\tilde{x})b_L^\dag(\tilde{x})\right]|\mathrm{vac}, 1\rangle+\nonumber\\
&&e_2|\mathrm{vac}, 2\rangle+e_3|\mathrm{vac},3\rangle,
\label{eigensate}
\end{eqnarray}
where $e_{2(3)}$ is the atomic excitation amplitude of the state $|2(3)\rangle$. According to the interaction $H_i$, we plot Fig. \ref{scattering} to describe the scattering with ansatz. The atom at each coupling point acts as an individual $\delta$-potential and thus the scattering can be equivalently treated as that in a series of cascaded small atoms in Refs. \cite{Kockum2014, Gough2009}. Therefore, the probability amplitude $\phi_{L(R)}^{\alpha(\beta)}(x)$  can be formally written as
$\phi_R^{\alpha}(x)=e^{i\alpha x}[\theta(x_1-x)+\sum_{n=1}^{N-1}t_n\theta(x-x_n)\theta(x_{n+1}-x)+
t_N\theta(x-x_N)]$,
$\phi_L^{\alpha}(x)=e^{-i\alpha x}[r_1\theta(x_1-x)+\sum_{n=2}^N r_n\theta(x-x_{n-1})\theta(x_n-x)]$,
$\phi_R^{\beta}(\tilde{x})=e^{i\beta \tilde{x}}[\sum_{m=1}^{M-1}\tilde{t}_m\theta(\tilde{x}-\tilde{x}_{m})\theta(\tilde{x}_{m+1}-\tilde{x})+\tilde{t}_M\theta(\tilde{x}-\tilde{x}_M)]$,
$\phi_L^{\beta}(\tilde{x})=e^{-i\beta \tilde{x}}[\tilde{r}_1\theta(\tilde{x}_1-\tilde{x})+\sum_{m=2}^M\tilde{r}_m\theta(\tilde{x}-\tilde{x}_{m-1})\theta(\tilde{x}_{m}-\tilde{x})]$,
where the Heaviside step function $\theta(x)$ is used to distinguish different scattering intervals. The joint transmission and reflection amplitudes in the intervals of waveguide $A(B)$ are denoted as $t_n (\tilde{t}_m)$ and $r_n(\tilde{r}_m)$, respectively. Note that these amplitudes represent the overall scatterings under steady states, which include higher-order scatterings resulting from multiple reflections between every two neighbor coupling points.
\begin{figure}
\centering
\includegraphics[width=0.46\textwidth]{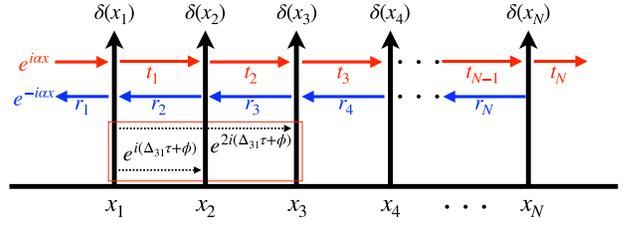}
\caption{Schematic of the scattering for the probe field. In real space, the giant atom behaves as a $\delta$ potential at each coupling point $x_n$. The joint transmission amplitudes $t_1, t_2, ..., t_{N-1}$ and reflection amplitudes $r_2, r_3, ...., r_N$ describe the single transmission and reflection between the neighboring two points, where the transmission amplitude $t_N$ and reflection amplitude $r_1$ contains all the accumulated phases. An example of the accumulated phase for the right-forward propagating process is given in the red box.}
\label{scattering}
\end{figure}

Substituting Eqs.~\eqref{Heff}-\eqref{eigensate} into stationary Schr\"odinger equation $H_{\mathrm{eff}}|\Psi\rangle=\omega|\Psi\rangle$, one can obtain the transmission amplitude of the probe field
\begin{equation}
t_N=\frac{(\Delta_F+i\gamma)(\Delta_S+\frac{i\gamma_3}{2})-|\Omega_c|^2}{(\Delta_F+i\gamma)(\Delta_S+\frac{i\gamma_3}{2}+i\Gamma_{31}^{(N)})-|\Omega_c|^2}
\label{transmission}
\end{equation}
and the excitation of the atom
\begin{equation}
e_3=\frac{(\Delta_F+i\gamma)\tilde{g}_{31}\sum_{n=1}^Ne^{(N-n)i(\Delta_{31}+\omega_3)\tau}}{(\Delta_F+i\gamma)(\Delta_S+\frac{i\gamma_3}{2}+i\Gamma_{31}^{(N)})-|\Omega_c|^2}
\label{Excitation}
\end{equation}
with $\Delta_F=\Delta_{31}-\Delta_{32}-\Delta_r^{(M)}$, $\gamma=\frac{i\gamma_2}{2}+i\Gamma_{21}^{(M)}$, $\Delta_S=\Delta_{31}-\Delta_{L}^{(N)}$, the detunings $\Delta_{31}=\omega-\omega_{31}$ and $\Delta_{32}=\nu_c-(\omega_{31}-\omega_{21})$. Both of them are affected by the frequency shift
\begin{equation}\label{LambR}
\Delta_r^{(M)}=\Gamma_{21}\sum_{m=1}^M(M-m)\sin(m\omega_{\beta} \tilde{\tau})
\end{equation}
and the effective decay rate
\begin{equation}\label{GammaR}
\Gamma_{21}^{(M)}=\Gamma_{21}\sum_{m=1}^M{\big [}\frac{M}{2}+(M-m)\cos(m\omega_{\beta}\tilde{\tau}){\big ]}
\end{equation}
induced by the waveguide $B$, where $\tilde{\tau}=\tilde{d}/\tilde{v}_g$ is a time delay and $\Gamma_{21}=2\tilde{g}_{21}^2/\tilde{v}_g$. Interestingly, in the case of $M=2$, the decay rate $\Gamma_{21}^{(2)}$ can be zero when $\omega_\beta\tilde\tau=(2k+1)\pi, k\in \mathbb{Z}^+$. This means that the spontaneous emission $|2\rangle\leftrightarrow|1\rangle$ can be totally eliminated, if we properly engineer the time delay $\tilde\tau$ with respect to a reservoir frequency $\omega_\beta$. Compared to traditional methods, e.g., employing an qubit with multiple Josephson junctions \cite{Manucharyan2013, Abdumalikov2010}, unique 3D cavities \cite{ Novikov2016}, or nested polariton states \cite{Long2018}, our elimination is realized by properly allocating couplings using multiple capacitors or piezoelectric transducers, which simplifies the systems. Hereafter, we assume that this elimination holds. In this way, we transform a $\Delta$-type atom to a $\Lambda$-type one.

In addition, the couplings to the waveguide $A$ induce the frequency-dependent Lamb shift
\begin{equation}
\Delta_L^{(N)}=\Gamma_{31}\sum_{n=1}^N(N-n)\sin\left(n\Delta_{31}\tau+n\phi\right)
\label{LambN}
\end{equation}
and the modified decay rate
\begin{equation}
\Gamma_{31}^{(N)}=\Gamma_{31}\sum_{n=1}^N{\Big[}\frac{N}{2}+(N-n)\cos\left(n\Delta_{31}\tau+n\phi\right){\Big]}
\label{GammaN}
\end{equation}
with $\Gamma_{31}=2\tilde{g}_{31}^2/v_g$, $\tau=d/v_g$, and $\phi=\omega_{31}\tau$. The first term in the addend of  Eq. \eqref{GammaN} denotes the joint decay rates attributed by the $N$ individual coupling points, while the second term is led by the spatial non-local couplings. Note that when $N=2$ and $\Omega_c=0$, the transmission amplitude $t_N$ \eqref{transmission} reduces to that of a two-level giant atom, which is consistent with Refs. \cite{Guo2017, Andersson2019}.
Also, the transmission amplitude $t_N$ \eqref{transmission} is modified by Eqs. \eqref{LambN} and \eqref{GammaN} with sinusoidals, which indicates that the transmission spectra would be different from those in traditional EIT. Particularly, when $N=1$, the sinusoidals vanishes such that the decay rate $\Gamma_{31}^{(N)}$ reduces to a constant $\Gamma_{31}/2$; i.e., the giant atom reduces to a small atom, which is consistent with that in Refs. \cite{Shen2009, Zhu2019}.

\begin{figure}
\centering
\includegraphics[width=0.46\textwidth]{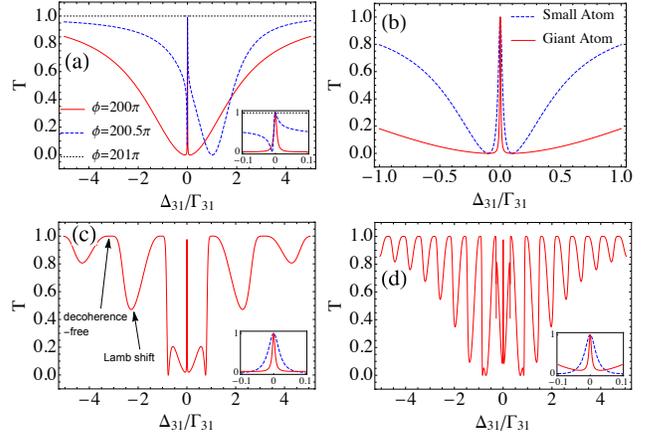}
\caption{Transmission spectra of the giant atom in the EIT regime with $N=2$ and $\Omega_c=0.1\Gamma_{31}$. (a) With a small $\tau=0.05/\Gamma_{31}$, the spectra have only one peak since where both the Lamb shift $\Delta_{31}^{(2)}$ and decay rate $\Gamma_{31}^{(2)}$ are mainly determined by the phase $\phi$. (b) Compared to small atoms, the transparent window $\Gamma_{w}\simeq|\Omega_c|^2/\Gamma_{31}^{(2)}$ is clearly modified by the phase $\phi$, where $\phi=200\pi,\tau=0.05/\Gamma_{31}$. In (c) $\tau=3/\Gamma_{31}$ and (d) $\tau=10/\Gamma_{31}$, we show the unconventional EIT spectra with increased time delays $\tau$ for a fixed phase $\phi=200\pi$. In both cases, $\Delta_{31}\tau$ dominate the variations of $\Delta_{31}^{(2)}$ and $\Gamma_{31}^{(2)}$ such that absorption peaks (arrow to Lamb shift) and transparent subpeaks (arrow to decoherence free) appear alternately. The blue dashed lines in insets represent the small atom case. Other parameters are: $\Delta_{32}=\gamma_2=\gamma_3=0$.}
\label{LambEIT}
\end{figure}

\noindent{\textbf{Unconventional EIT---}}  With the above results, we discover the following unconventional EIT. We let $N=2$ and the time delay $\tau=\{0.05, 3, 10\}/\Gamma_{31}$  cover both the microwave-coupling and acoustic-coupling regimes in Refs. \cite{Vadiraj2021, Andersson2019}.

In the EIT regime, we plot the transmission spectra $T=|t_2|^2$ in Fig. \ref{LambEIT}.
When the time delay $\tau=0.05/\Gamma_{31}$ is sufficiently small and $\Delta_{31}\ll\omega_{31}$, the Lamb shift $\Delta_{L}^{(2)}$ and the decay rate $\Gamma_{31}^{(2)}$ are mainly determined by the phase $\phi$ such that the spectrum has a single absorption peak determined by the frequency $\omega_{31}$ as shown in Fig. \ref{LambEIT}(a), which is consistent with that in Ref. \cite{Kockum2014}. Particularly, when $\phi=(2k+1)\pi$, we have $\Delta_{L}^{(2)}=\Gamma_{31}^{(2)}=0$ where the system is decoherence-free; i.e., the giant atom is totally decoupled from the waveguide $A$, although there still exists a non-zero coupling $\Gamma_{31}$. This is represented by the black dot line in Fig. \ref{LambEIT}(a). Further, with a given phase $\phi=200\pi$, we increase the time delay $\tau$ so that decoherence-free bands and Lamb-shift-induced absorption peaks appear alternatively and become dense, as shown in Figs. \ref{LambEIT}(b)-(d). This is because the absorption peaks with a modified linewidth $\Gamma_{31}^{(2)}$ appear in the single-photon resonance $\Delta_{31}=\Delta_{L}^{(2)}$ (see $\Delta_S$ in Eq. \eqref{transmission}). In addition, since the induced transparency occurs at two-photon resonance $\Delta_{31}-\Delta_{32}=0$ (see $\Delta_F$ in Eq. \eqref{transmission}), the transparency window $\Gamma_w\simeq|\Omega_c|^2/\Gamma_{31}^{(2)}$ is only modified by the decay rate $\Gamma_{31}^{(2)}$ which depends on the initial phase $\phi$, as shown in Fig. \ref{LambEIT}(b) and the insets in Figs. \ref{LambEIT}(c)-(d).
Therefore, our results predict that unconventional EIT can be observed when the time delay between two neighboring coupling points is comparable to the relaxation time $1/\Gamma_{31}$.

Note that by increasing the Rabi frequency $\Omega_c$, similar solutions can also be found in the Autler-Townes Splitting (ATS) regime \cite{Autler1955}. Besides, a stronger modification can be obtained by increasing the number of the coupling points. For instance, the maximum value of $\Gamma_{31}^{(3)}=2\Gamma_{31}\left[\frac{1}{2}+\cos(\Delta_{31}\tau+\phi)\right]^2$ is greatly enhanced compared to $\Gamma_{31}^{(2)}=2\Gamma_{31}\cos^2\left(\frac{\Delta_{31}\tau+\phi}{2}\right)$.
Also, the sinusoidal form of Eqs. \eqref{LambN}-\eqref{GammaN} indicates that it results from a boundary condition of standing waves, i.e., the interference between bidirectional propagating waves form standing waves inside each scattering interval. Details of the above points can be found in the supplemental  material \cite{SM2020}. In addition, this tunable window may provide enhancement of  sensitivity in qubit-assisted quantum sensing \cite{Fleischhauer2000, Meyer2021}.

\noindent{\textbf{Time-delayed master equation---}} Actually, this unconventional EIT is a non-Markovian effect in the giant atom. Here, we firstly present a time-delayed master equation
\begin{equation}
\dot{\rho}_a(t)=\mathcal{L}_{loc}\rho_a(t)+\sum_{n=1}^N\mathcal{L}'_{D}\rho_a(t-n\tau)+\sum_{m=1}^M\mathcal{L}''_{D}\rho_a(t-m\tilde{\tau}),
\label{NonMarkovianEquation}
\end{equation}
where the local evolution $\mathcal{L}_{loc}\rho_a(t)=(\mathcal{L}_0+\mathcal{L}_{D})\rho_a(t)$ includes a free evolution $\mathcal{L}_0\rho_a(t)=-i[V_c(t),\rho_a(t)]$ with
$V_c(t)=(\sum_{n=1}^N\Omega_pe^{-i\Delta_{31}t+i\tilde\alpha x_n}|3\rangle\langle1|+\Omega_ce^{-i\Delta_{32}t}|3\rangle\langle2|+H.c.)$ and a dissipative term $\mathcal{L}_{loc}\rho_a(t)=(\gamma_{31}+N\Gamma_{31})\mathcal{D}[|1\rangle\langle3|]\rho_a(t)+
(\gamma_{32}+M\Gamma_{32})\mathcal{D}[|2\rangle\langle3]\rho_a(t)+\gamma_{2}^{\phi}\mathcal{D}[|2\rangle\langle2|]\rho_a(t)+
\gamma_{3}^{\phi}\mathcal{D}[|3\rangle\langle3|]\rho_a(t)+2\Gamma_{21}\mathcal{D}[|1\rangle\langle2|]\rho_a(t)$
Here, $\Omega_p$, $\gamma_{31(2)}$ and $\gamma_{2(3)}^{\phi}$ represent the amplitude of a weak probe field, the non-waveguide decay rate, and the pure dephasing rate, respectively, and $\mathcal{D}[O]\rho(t)$ denotes the Lindblad superoperator for an arbitrary operator $O$.

Different from conventional master equations for giant atoms, we find non-locality resulted from distant multiple couplings attributes the time nonlocal terms
\begin{eqnarray}
\mathcal{L}'_{D}\rho_a(t-n\tau)&&=\mathcal{L}'_0\rho_a(t-n\tau)+2\Gamma'_{31}\mathcal{D}[|1\rangle\langle3|]\rho_a(t-n\tau),\nonumber\\
\mathcal{L}''_{D}\rho_a(t-m\tilde{\tau})&&=\mathcal{L}''_0\rho_a(t-m\tilde{\tau})+
2\Gamma''_{21}\mathcal{D}[|1\rangle\langle2|]\rho_a(t-m\tilde{\tau}),\nonumber\\\label{Lindbladian1}
\end{eqnarray}
with $\mathcal{L}'_0\rho_a(t-n\tau)=-i\Delta'_L[|3\rangle\langle3|,\rho_a(t-n\tau)]$ and
$\mathcal{L}''_0\rho_a(t-m\tilde{\tau})=-i\Delta''_r[|2\rangle\langle2|,\rho_a(t-m\tilde{\tau})]$.
The damping rates $\Gamma'_{31}=\Gamma_{31}(N-n)\cos(n\omega_{31}\tau)$ and
$\Gamma''_{21}=\Gamma_{21}(M-m)\cos(\omega_{21}\tilde{\tau})$ and the detunings $\Delta'_{L}=\Gamma_{31}(N-n)\sin(n\omega_{31}\tau)$ and $\Delta''_{r}=\Gamma_{21}(M-m)\sin(\omega_{21}\tilde{\tau})$ are in a sinusoidal form.
Note that we do not use the single-excitation assumption in the derivation and hence it can be extended to future general cases.

The steady states of Eq. \eqref{NonMarkovianEquation} reads
\begin{equation}
\tilde{\rho}_{31}=\frac{1}{2\pi}\frac{\left(\Delta_F+i\gamma\right)\Omega_p\sum_{n=1}^Ne^{(N-n)i(\Delta_{31}\tau+\phi)}}{(\Delta_F+i\gamma)(\Delta_S+\frac{i\gamma_3}{2}+i\Gamma_{31}^{(N)})-|\Omega_c|^2},
\end{equation}
which is consistent with the excitation $e_3$ \eqref{Excitation} under the replacement $\tilde{g}_{31}\rightarrow\Omega_p$. The time delay induced by the spatial non-locality is embedded in the sinusoidals in Eqs. \eqref{LambN}-\eqref{GammaN}  which allows a strictly negative decay rate for some detuning $\Delta_{31}$. This means that the information can be transferred back to the system from the environment \cite{Breuer2009, Laine2010, Hall2014}, i.e., the non-Markovian effect occurs. It results from the inherent non-localities rather than the spectral property of interacting environments for common non-Markovian dynamics \cite{Breuer2016, Xue2012, Zhang2012}.

\noindent{\textbf{Further discussions---}}In quantum optics, EIT is commonly described by a semi-classical master equation \cite{Fleischhauer2005} which is not a suitable description for giant atoms. This is because the multi-peak spectra and non-Markovian effects actually results from the Lindbladians \eqref{Lindbladian1} which come from a quantized waveguide-atom Hamiltonian. This also explains the discrepancy from the data and the fit in Fig. 8 of Ref. \cite{Vadiraj2021}. A detailed comparison between semi-classical and full quantum theories are shown in Sec. VII of the supplimentary\cite{SM2020}.

On the other hand, although similar systems with non-locality appear in the existing works, non-locality therein has been ruined by improper Markovian approximation. For instance, with the replacements $\rho_a(t-n\tau) \rightarrow\rho_{a}(t)$ in master equations or $e^{i\alpha x}\rightarrow e^{i\omega_{31}x/v_g}$ in the real-space method, the modified decay rate $\Gamma_{31}^{(N)}$ only depends on the atomic frequency $\omega_{31}$ \cite{Kockum2014, Fang2015, Calajo2019}.

In addition, giant atoms are defined as an artificial atom whose size is much larger than the wavelength of its interacting field, or whose relaxation time is much less than the propagating time of the field to pass the entire atom \cite{Delsing2019}. However, it is improper for Ref. \cite{Andersson2020}, where the atom is coupled to acoustics with one transducer and the standard EIT is observed only. Therefore, we update that the giant atoms should be defined as that atoms are multi-point coupled to the fields.

\noindent{\textbf{Conclusion---}} In summary, we have firstly discovered unconventional EIT and ATS in a single giant atom using our consistent theory including a real-space method and a time delayed master equation. This interesting phenomenon is a typical quantum effect which is not observable semi-classically\cite{Andersson2020, Vadiraj2021}. Our theory indicates that to experimentally observe it the time delay between two neighboring coupling points must be comparable to the relaxation time and this is why it has not been experimentally observed so far. Also, the physics behind this phenomenon is the inherent non-locality of giant atoms which leads to the standing waves in space and retardations in time.

Our work has established a unified framework for exploring possible applications of nonlocal systems in quantum filters \cite{Xue2017A, Xue2017B, Xue2020}, switches \cite{Hoi2011, Zhu2019}, memories \cite{Afzelius2015}, slowing/stopping light \cite{Kasapi1995, Xiao1995, Schmidt1996}, and lasers without inversion \cite{Scully1989, Fry1993, Nottelmann1993}. Especially, one possible direction is to apply the unconventional EIT in a long-time storage of quantum information so as to achieve long-range quantum communication.

\noindent{\textbf{Acknowledgements---}} We thank W. Z. Jia, Z. H. Peng, Q. Y. Cai, L. Du, F. Ciccarello, and X. M. Jin for fruitful discussion. This work is supported by the National Natural Science Foundation of China (NSFC) under Grants 61873162, 61973317. This work was also supported by the Open Research Project of the State Key Laboratory of Industrial Control Technology, Zhejiang University, China (No. ICT2021B24).

\bibliography{Letter-11-30}

\providecommand{\noopsort}[1]{}\providecommand{\singleletter}[1]{#1}%
\begin{thebibliography}{52}%
\makeatletter
\providecommand \@ifxundefined [1]{%
 \@ifx{#1\undefined}
}%
\providecommand \@ifnum [1]{%
 \ifnum #1\expandafter \@firstoftwo
 \else \expandafter \@secondoftwo
 \fi
}%
\providecommand \@ifx [1]{%
 \ifx #1\expandafter \@firstoftwo
 \else \expandafter \@secondoftwo
 \fi
}%
\providecommand \natexlab [1]{#1}%
\providecommand \enquote  [1]{``#1''}%
\providecommand \bibnamefont  [1]{#1}%
\providecommand \bibfnamefont [1]{#1}%
\providecommand \citenamefont [1]{#1}%
\providecommand \href@noop [0]{\@secondoftwo}%
\providecommand \href [0]{\begingroup \@sanitize@url \@href}%
\providecommand \@href[1]{\@@startlink{#1}\@@href}%
\providecommand \@@href[1]{\endgroup#1\@@endlink}%
\providecommand \@sanitize@url [0]{\catcode `\\12\catcode `\$12\catcode
  `\&12\catcode `\#12\catcode `\^12\catcode `\_12\catcode `\%12\relax}%
\providecommand \@@startlink[1]{}%
\providecommand \@@endlink[0]{}%
\providecommand \url  [0]{\begingroup\@sanitize@url \@url }%
\providecommand \@url [1]{\endgroup\@href {#1}{\urlprefix }}%
\providecommand \urlprefix  [0]{URL }%
\providecommand \Eprint [0]{\href }%
\providecommand \doibase [0]{https://doi.org/}%
\providecommand \selectlanguage [0]{\@gobble}%
\providecommand \bibinfo  [0]{\@secondoftwo}%
\providecommand \bibfield  [0]{\@secondoftwo}%
\providecommand \translation [1]{[#1]}%
\providecommand \BibitemOpen [0]{}%
\providecommand \bibitemStop [0]{}%
\providecommand \bibitemNoStop [0]{.\EOS\space}%
\providecommand \EOS [0]{\spacefactor3000\relax}%
\providecommand \BibitemShut  [1]{\csname bibitem#1\endcsname}%
\let\auto@bib@innerbib\@empty
\bibitem [{\citenamefont {Gustafsson}\ \emph {et~al.}(2014)\citenamefont
  {Gustafsson}, \citenamefont {Aref}, \citenamefont {Kockum}, \citenamefont
  {Ekstr\"{o}m}, \citenamefont {Johansson},\ and\ \citenamefont
  {Delsing}}]{Gustafsson2014}%
  \BibitemOpen
  \bibfield  {author} {\bibinfo {author} {\bibfnamefont {M.~V.}\ \bibnamefont
  {Gustafsson}}, \bibinfo {author} {\bibfnamefont {T.}~\bibnamefont {Aref}},
  \bibinfo {author} {\bibfnamefont {A.~F.}\ \bibnamefont {Kockum}}, \bibinfo
  {author} {\bibfnamefont {M.~K.}\ \bibnamefont {Ekstr\"{o}m}}, \bibinfo
  {author} {\bibfnamefont {G.}~\bibnamefont {Johansson}},\ and\ \bibinfo
  {author} {\bibfnamefont {P.}~\bibnamefont {Delsing}},\ }\bibfield  {title}
  {\bibinfo {title} {Propagating phonons coupled to an artificial atom},\
  }\href {https://doi.org/10.1126/science.1257219} {\bibfield  {journal}
  {\bibinfo  {journal} {Science}\ }\textbf {\bibinfo {volume} {346}},\ \bibinfo
  {pages} {207} (\bibinfo {year} {2014})}\BibitemShut {NoStop}%
\bibitem [{\citenamefont {Schuetz}\ \emph {et~al.}(2015)\citenamefont
  {Schuetz}, \citenamefont {Kessler}, \citenamefont {Giedke}, \citenamefont
  {Vandersypen}, \citenamefont {Lukin},\ and\ \citenamefont
  {Cirac}}]{Schuetz2015}%
  \BibitemOpen
  \bibfield  {author} {\bibinfo {author} {\bibfnamefont {M.~J.~A.}\
  \bibnamefont {Schuetz}}, \bibinfo {author} {\bibfnamefont {E.~M.}\
  \bibnamefont {Kessler}}, \bibinfo {author} {\bibfnamefont {G.}~\bibnamefont
  {Giedke}}, \bibinfo {author} {\bibfnamefont {L.~M.~K.}\ \bibnamefont
  {Vandersypen}}, \bibinfo {author} {\bibfnamefont {M.~D.}\ \bibnamefont
  {Lukin}},\ and\ \bibinfo {author} {\bibfnamefont {J.~I.}\ \bibnamefont
  {Cirac}},\ }\bibfield  {title} {\bibinfo {title} {{Universal quantum
  transducers based on surface acoustic waves}},\ }\href
  {https://doi.org/10.1103/physrevx.5.031031} {\bibfield  {journal} {\bibinfo
  {journal} {Phys. Rev. X}\ }\textbf {\bibinfo {volume} {5}},\ \bibinfo {pages}
  {031031} (\bibinfo {year} {2015})}\BibitemShut {NoStop}%
\bibitem [{\citenamefont {Chu}\ \emph {et~al.}(2017)\citenamefont {Chu},
  \citenamefont {Kharel}, \citenamefont {Renninger}, \citenamefont {Burkhart},
  \citenamefont {Frunzio}, \citenamefont {Rakich},\ and\ \citenamefont
  {Schoelkopf}}]{Chu2017}%
  \BibitemOpen
  \bibfield  {author} {\bibinfo {author} {\bibfnamefont {Y.}~\bibnamefont
  {Chu}}, \bibinfo {author} {\bibfnamefont {P.}~\bibnamefont {Kharel}},
  \bibinfo {author} {\bibfnamefont {W.~H.}\ \bibnamefont {Renninger}}, \bibinfo
  {author} {\bibfnamefont {L.~D.}\ \bibnamefont {Burkhart}}, \bibinfo {author}
  {\bibfnamefont {L.}~\bibnamefont {Frunzio}}, \bibinfo {author} {\bibfnamefont
  {P.~T.}\ \bibnamefont {Rakich}},\ and\ \bibinfo {author} {\bibfnamefont
  {R.~J.}\ \bibnamefont {Schoelkopf}},\ }\bibfield  {title} {\bibinfo {title}
  {Quantum acoustics with superconducting qubits},\ }\href
  {https://science.sciencemag.org/content/358/6360/199} {\bibfield  {journal}
  {\bibinfo  {journal} {Science}\ }\textbf {\bibinfo {volume} {358}},\ \bibinfo
  {pages} {199} (\bibinfo {year} {2017})}\BibitemShut {NoStop}%
\bibitem [{\citenamefont {Chu}\ \emph {et~al.}(2018)\citenamefont {Chu},
  \citenamefont {Kharel}, \citenamefont {Yoon}, \citenamefont {Frunzio},
  \citenamefont {Rakich},\ and\ \citenamefont {Schoelkopf}}]{Chu2018}%
  \BibitemOpen
  \bibfield  {author} {\bibinfo {author} {\bibfnamefont {Y.}~\bibnamefont
  {Chu}}, \bibinfo {author} {\bibfnamefont {P.}~\bibnamefont {Kharel}},
  \bibinfo {author} {\bibfnamefont {T.}~\bibnamefont {Yoon}}, \bibinfo {author}
  {\bibfnamefont {L.}~\bibnamefont {Frunzio}}, \bibinfo {author} {\bibfnamefont
  {P.~T.}\ \bibnamefont {Rakich}},\ and\ \bibinfo {author} {\bibfnamefont
  {R.~J.}\ \bibnamefont {Schoelkopf}},\ }\bibfield  {title} {\bibinfo {title}
  {Creation and control of multi-phonon fock states in a bulk acoustic-wave
  resonator},\ }\href {https://doi.org/10.1038/s41586-018-0717-7} {\bibfield
  {journal} {\bibinfo  {journal} {Nature}\ }\textbf {\bibinfo {volume} {563}},\
  \bibinfo {pages} {666} (\bibinfo {year} {2018})}\BibitemShut {NoStop}%
\bibitem [{\citenamefont {Satzinger}\ \emph {et~al.}(2018)\citenamefont
  {Satzinger}, \citenamefont {Zhong}, \citenamefont {Chang}, \citenamefont
  {Peairs}, \citenamefont {Bienfait}, \citenamefont {Chou}, \citenamefont
  {Cleland}, \citenamefont {Conner}, \citenamefont {Dumur}, \citenamefont
  {Grebel}, \citenamefont {Gutierrez}, \citenamefont {November}, \citenamefont
  {Povey}, \citenamefont {Whiteley}, \citenamefont {Awschalom}, \citenamefont
  {Schuster},\ and\ \citenamefont {Cleland}}]{Satzinger2018}%
  \BibitemOpen
  \bibfield  {author} {\bibinfo {author} {\bibfnamefont {K.~J.}\ \bibnamefont
  {Satzinger}}, \bibinfo {author} {\bibfnamefont {Y.~P.}\ \bibnamefont
  {Zhong}}, \bibinfo {author} {\bibfnamefont {H.-S.}\ \bibnamefont {Chang}},
  \bibinfo {author} {\bibfnamefont {G.~A.}\ \bibnamefont {Peairs}}, \bibinfo
  {author} {\bibfnamefont {A.}~\bibnamefont {Bienfait}}, \bibinfo {author}
  {\bibfnamefont {M.-H.}\ \bibnamefont {Chou}}, \bibinfo {author}
  {\bibfnamefont {A.~Y.}\ \bibnamefont {Cleland}}, \bibinfo {author}
  {\bibfnamefont {C.~R.}\ \bibnamefont {Conner}}, \bibinfo {author}
  {\bibfnamefont {E.}~\bibnamefont {Dumur}}, \bibinfo {author} {\bibfnamefont
  {J.}~\bibnamefont {Grebel}}, \bibinfo {author} {\bibfnamefont
  {I.}~\bibnamefont {Gutierrez}}, \bibinfo {author} {\bibfnamefont {B.~H.}\
  \bibnamefont {November}}, \bibinfo {author} {\bibfnamefont {R.~G.}\
  \bibnamefont {Povey}}, \bibinfo {author} {\bibfnamefont {S.~J.}\ \bibnamefont
  {Whiteley}}, \bibinfo {author} {\bibfnamefont {D.~D.}\ \bibnamefont
  {Awschalom}}, \bibinfo {author} {\bibfnamefont {D.~I.}\ \bibnamefont
  {Schuster}},\ and\ \bibinfo {author} {\bibfnamefont {A.~N.}\ \bibnamefont
  {Cleland}},\ }\bibfield  {title} {\bibinfo {title} {Quantum control of
  surface acoustic-wave phonons},\ }\href
  {https://doi.org/10.1038/s41586-018-0719-5} {\bibfield  {journal} {\bibinfo
  {journal} {Nature}\ }\textbf {\bibinfo {volume} {563}},\ \bibinfo {pages}
  {661} (\bibinfo {year} {2018})}\BibitemShut {NoStop}%
\bibitem [{\citenamefont {Ekstr\"om}\ \emph {et~al.}(2019)\citenamefont
  {Ekstr\"om}, \citenamefont {Aref}, \citenamefont {Ask}, \citenamefont
  {Andersson}, \citenamefont {Suri}, \citenamefont {Sanada}, \citenamefont
  {Johansson},\ and\ \citenamefont {Delsing}}]{Ekstrom2019}%
  \BibitemOpen
  \bibfield  {author} {\bibinfo {author} {\bibfnamefont {M.~K.}\ \bibnamefont
  {Ekstr\"om}}, \bibinfo {author} {\bibfnamefont {T.}~\bibnamefont {Aref}},
  \bibinfo {author} {\bibfnamefont {A.}~\bibnamefont {Ask}}, \bibinfo {author}
  {\bibfnamefont {G.}~\bibnamefont {Andersson}}, \bibinfo {author}
  {\bibfnamefont {B.}~\bibnamefont {Suri}}, \bibinfo {author} {\bibfnamefont
  {H.}~\bibnamefont {Sanada}}, \bibinfo {author} {\bibfnamefont
  {G.}~\bibnamefont {Johansson}},\ and\ \bibinfo {author} {\bibfnamefont
  {P.}~\bibnamefont {Delsing}},\ }\bibfield  {title} {\bibinfo {title} {Towards
  phonon routing: controlling propagating acoustic waves in the quantum
  regime},\ }\href {https://doi.org/10.1088/1367-2630/ab5ca5} {\bibfield
  {journal} {\bibinfo  {journal} {New J. Phys.}\ }\textbf {\bibinfo {volume}
  {21}},\ \bibinfo {pages} {123013} (\bibinfo {year} {2019})}\BibitemShut
  {NoStop}%
\bibitem [{\citenamefont {Manenti}\ \emph {et~al.}(2017)\citenamefont
  {Manenti}, \citenamefont {Kockum}, \citenamefont {Patterson}, \citenamefont
  {Behrle}, \citenamefont {Rahamim}, \citenamefont {Tancredi}, \citenamefont
  {Nori},\ and\ \citenamefont {Leek}}]{Manenti2017}%
  \BibitemOpen
  \bibfield  {author} {\bibinfo {author} {\bibfnamefont {R.}~\bibnamefont
  {Manenti}}, \bibinfo {author} {\bibfnamefont {A.~F.}\ \bibnamefont {Kockum}},
  \bibinfo {author} {\bibfnamefont {A.}~\bibnamefont {Patterson}}, \bibinfo
  {author} {\bibfnamefont {T.}~\bibnamefont {Behrle}}, \bibinfo {author}
  {\bibfnamefont {J.}~\bibnamefont {Rahamim}}, \bibinfo {author} {\bibfnamefont
  {G.}~\bibnamefont {Tancredi}}, \bibinfo {author} {\bibfnamefont
  {F.}~\bibnamefont {Nori}},\ and\ \bibinfo {author} {\bibfnamefont {P.~J.}\
  \bibnamefont {Leek}},\ }\bibfield  {title} {\bibinfo {title} {Circuit quantum
  acoustodynamics with surface acoustic waves},\ }\href
  {https://doi.org/10.1038/s41467-017-01063-9} {\bibfield  {journal} {\bibinfo
  {journal} {Nat. Comm.}\ }\textbf {\bibinfo {volume} {8}},\ \bibinfo {pages}
  {975} (\bibinfo {year} {2017})}\BibitemShut {NoStop}%
\bibitem [{\citenamefont {Andersson}\ \emph {et~al.}(2019)\citenamefont
  {Andersson}, \citenamefont {Suri}, \citenamefont {Guo}, \citenamefont
  {Aref},\ and\ \citenamefont {Delsing}}]{Andersson2019}%
  \BibitemOpen
  \bibfield  {author} {\bibinfo {author} {\bibfnamefont {G.}~\bibnamefont
  {Andersson}}, \bibinfo {author} {\bibfnamefont {B.}~\bibnamefont {Suri}},
  \bibinfo {author} {\bibfnamefont {L.}~\bibnamefont {Guo}}, \bibinfo {author}
  {\bibfnamefont {T.}~\bibnamefont {Aref}},\ and\ \bibinfo {author}
  {\bibfnamefont {P.}~\bibnamefont {Delsing}},\ }\bibfield  {title} {\bibinfo
  {title} {Non-exponential decay of a giant artificial atom},\ }\href
  {https://doi.org/10.1038/s41567-019-0605-6} {\bibfield  {journal} {\bibinfo
  {journal} {Nat. Phys.}\ }\textbf {\bibinfo {volume} {15}},\ \bibinfo {pages}
  {1123} (\bibinfo {year} {2019})}\BibitemShut {NoStop}%
\bibitem [{\citenamefont {Kannan}\ \emph {et~al.}(2020)\citenamefont {Kannan},
  \citenamefont {Ruckriegel}, \citenamefont {Campbell}, \citenamefont
  {Frisk~Kockum}, \citenamefont {Braumuller}, \citenamefont {Kim},
  \citenamefont {Kjaergaard}, \citenamefont {Krantz}, \citenamefont {Melville},
  \citenamefont {Niedzielski}, \citenamefont {Vepsalainen}, \citenamefont
  {Winik}, \citenamefont {Yoder}, \citenamefont {Nori}, \citenamefont
  {Orlando}, \citenamefont {Gustavsson},\ and\ \citenamefont
  {Oliver}}]{Kannan2020}%
  \BibitemOpen
  \bibfield  {author} {\bibinfo {author} {\bibfnamefont {B.}~\bibnamefont
  {Kannan}}, \bibinfo {author} {\bibfnamefont {M.~J.}\ \bibnamefont
  {Ruckriegel}}, \bibinfo {author} {\bibfnamefont {D.~L.}\ \bibnamefont
  {Campbell}}, \bibinfo {author} {\bibfnamefont {A.}~\bibnamefont
  {Frisk~Kockum}}, \bibinfo {author} {\bibfnamefont {J.}~\bibnamefont
  {Braumuller}}, \bibinfo {author} {\bibfnamefont {D.~K.}\ \bibnamefont {Kim}},
  \bibinfo {author} {\bibfnamefont {M.}~\bibnamefont {Kjaergaard}}, \bibinfo
  {author} {\bibfnamefont {P.}~\bibnamefont {Krantz}}, \bibinfo {author}
  {\bibfnamefont {A.}~\bibnamefont {Melville}}, \bibinfo {author}
  {\bibfnamefont {B.~M.}\ \bibnamefont {Niedzielski}}, \bibinfo {author}
  {\bibfnamefont {A.}~\bibnamefont {Vepsalainen}}, \bibinfo {author}
  {\bibfnamefont {R.}~\bibnamefont {Winik}}, \bibinfo {author} {\bibfnamefont
  {J.~L.}\ \bibnamefont {Yoder}}, \bibinfo {author} {\bibfnamefont
  {F.}~\bibnamefont {Nori}}, \bibinfo {author} {\bibfnamefont {T.~P.}\
  \bibnamefont {Orlando}}, \bibinfo {author} {\bibfnamefont {S.}~\bibnamefont
  {Gustavsson}},\ and\ \bibinfo {author} {\bibfnamefont {W.~D.}\ \bibnamefont
  {Oliver}},\ }\bibfield  {title} {\bibinfo {title} {Waveguide quantum
  electrodynamics with superconducting artificial giant atoms},\ }\href
  {https://doi.org/10.1038/s41586-020-2529-9} {\bibfield  {journal} {\bibinfo
  {journal} {Nature}\ }\textbf {\bibinfo {volume} {583}},\ \bibinfo {pages}
  {775} (\bibinfo {year} {2020})}\BibitemShut {NoStop}%
\bibitem [{\citenamefont {Frisk~Kockum}(2020)}]{Kockum2020}%
  \BibitemOpen
  \bibfield  {author} {\bibinfo {author} {\bibfnamefont {A.}~\bibnamefont
  {Frisk~Kockum}},\ }\bibinfo {title} {Quantum optics with giant atoms—the
  first five years},\ in\ \href {https://doi.org/10.1007/978-981-15-5191-8_12}
  {\emph {\bibinfo {booktitle} {Mathematics for Industry}}}\ (\bibinfo
  {publisher} {Springer Singapore},\ \bibinfo {year} {2020})\ p.\ \bibinfo
  {pages} {125–146}\BibitemShut {NoStop}%
\bibitem [{\citenamefont {Frisk~Kockum}\ \emph {et~al.}(2014)\citenamefont
  {Frisk~Kockum}, \citenamefont {Delsing},\ and\ \citenamefont
  {Johansson}}]{Kockum2014}%
  \BibitemOpen
  \bibfield  {author} {\bibinfo {author} {\bibfnamefont {A.}~\bibnamefont
  {Frisk~Kockum}}, \bibinfo {author} {\bibfnamefont {P.}~\bibnamefont
  {Delsing}},\ and\ \bibinfo {author} {\bibfnamefont {G.}~\bibnamefont
  {Johansson}},\ }\bibfield  {title} {\bibinfo {title} {Designing
  frequency-dependent relaxation rates and lamb shifts for a giant artificial
  atom},\ }\href {https://doi.org/10.1103/PhysRevA.90.013837} {\bibfield
  {journal} {\bibinfo  {journal} {Phys. Rev. A}\ }\textbf {\bibinfo {volume}
  {90}},\ \bibinfo {pages} {013837} (\bibinfo {year} {2014})}\BibitemShut
  {NoStop}%
\bibitem [{\citenamefont {Guo}\ \emph {et~al.}(2017)\citenamefont {Guo},
  \citenamefont {Grimsmo}, \citenamefont {Kockum}, \citenamefont {Pletyukhov},\
  and\ \citenamefont {Johansson}}]{Guo2017}%
  \BibitemOpen
  \bibfield  {author} {\bibinfo {author} {\bibfnamefont {L.}~\bibnamefont
  {Guo}}, \bibinfo {author} {\bibfnamefont {A.}~\bibnamefont {Grimsmo}},
  \bibinfo {author} {\bibfnamefont {A.~F.}\ \bibnamefont {Kockum}}, \bibinfo
  {author} {\bibfnamefont {M.}~\bibnamefont {Pletyukhov}},\ and\ \bibinfo
  {author} {\bibfnamefont {G.}~\bibnamefont {Johansson}},\ }\bibfield  {title}
  {\bibinfo {title} {Giant acoustic atom: A single quantum system with a
  deterministic time delay},\ }\href
  {https://doi.org/10.1103/PhysRevA.95.053821} {\bibfield  {journal} {\bibinfo
  {journal} {Phys. Rev. A}\ }\textbf {\bibinfo {volume} {95}},\ \bibinfo
  {pages} {053821} (\bibinfo {year} {2017})}\BibitemShut {NoStop}%
\bibitem [{\citenamefont {Guo}\ \emph {et~al.}(2020)\citenamefont {Guo},
  \citenamefont {Kockum}, \citenamefont {Marquardt},\ and\ \citenamefont
  {Johansson}}]{Guo2020}%
  \BibitemOpen
  \bibfield  {author} {\bibinfo {author} {\bibfnamefont {L.}~\bibnamefont
  {Guo}}, \bibinfo {author} {\bibfnamefont {A.~F.}\ \bibnamefont {Kockum}},
  \bibinfo {author} {\bibfnamefont {F.}~\bibnamefont {Marquardt}},\ and\
  \bibinfo {author} {\bibfnamefont {G.}~\bibnamefont {Johansson}},\ }\bibfield
  {title} {\bibinfo {title} {Oscillating bound states for a giant atom},\
  }\href {https://doi.org/10.1103/PhysRevResearch.2.043014} {\bibfield
  {journal} {\bibinfo  {journal} {Phys. Rev. Research}\ }\textbf {\bibinfo
  {volume} {2}},\ \bibinfo {pages} {043014} (\bibinfo {year}
  {2020})}\BibitemShut {NoStop}%
\bibitem [{\citenamefont {Kockum}\ \emph {et~al.}(2018)\citenamefont {Kockum},
  \citenamefont {Johansson},\ and\ \citenamefont {Nori}}]{Kockum2018}%
  \BibitemOpen
  \bibfield  {author} {\bibinfo {author} {\bibfnamefont {A.~F.}\ \bibnamefont
  {Kockum}}, \bibinfo {author} {\bibfnamefont {G.}~\bibnamefont {Johansson}},\
  and\ \bibinfo {author} {\bibfnamefont {F.}~\bibnamefont {Nori}},\ }\bibfield
  {title} {\bibinfo {title} {{Decoherence-free interaction between giant atoms
  in waveguide quantum electrodynamics}},\ }\href
  {https://doi.org/10.1103/physrevlett.120.140404} {\bibfield  {journal}
  {\bibinfo  {journal} {Phys. Rev. Lett.}\ }\textbf {\bibinfo {volume} {120}},\
  \bibinfo {pages} {140404} (\bibinfo {year} {2018})}\BibitemShut {NoStop}%
\bibitem [{\citenamefont {Harris}\ \emph {et~al.}(1990)\citenamefont {Harris},
  \citenamefont {Field},\ and\ \citenamefont {Imamoglu}}]{Harris1990}%
  \BibitemOpen
  \bibfield  {author} {\bibinfo {author} {\bibfnamefont {S.~E.}\ \bibnamefont
  {Harris}}, \bibinfo {author} {\bibfnamefont {J.~E.}\ \bibnamefont {Field}},\
  and\ \bibinfo {author} {\bibfnamefont {A.}~\bibnamefont {Imamoglu}},\
  }\bibfield  {title} {\bibinfo {title} {Nonlinear optical processes using
  electromagnetically induced transparency},\ }\href
  {https://doi.org/10.1103/PhysRevLett.64.1107} {\bibfield  {journal} {\bibinfo
   {journal} {Phys. Rev. Lett.}\ }\textbf {\bibinfo {volume} {64}},\ \bibinfo
  {pages} {1107} (\bibinfo {year} {1990})}\BibitemShut {NoStop}%
\bibitem [{\citenamefont {Boller}\ \emph {et~al.}(1991)\citenamefont {Boller},
  \citenamefont {Imamoglu},\ and\ \citenamefont {Harris}}]{Boller1991}%
  \BibitemOpen
  \bibfield  {author} {\bibinfo {author} {\bibfnamefont {K.-J.}\ \bibnamefont
  {Boller}}, \bibinfo {author} {\bibfnamefont {A.}~\bibnamefont {Imamoglu}},\
  and\ \bibinfo {author} {\bibfnamefont {S.~E.}\ \bibnamefont {Harris}},\
  }\bibfield  {title} {\bibinfo {title} {Observation of electromagnetically
  induced transparency},\ }\href {https://doi.org/10.1103/PhysRevLett.66.2593}
  {\bibfield  {journal} {\bibinfo  {journal} {Phys. Rev. Lett.}\ }\textbf
  {\bibinfo {volume} {66}},\ \bibinfo {pages} {2593} (\bibinfo {year}
  {1991})}\BibitemShut {NoStop}%
\bibitem [{\citenamefont {Fleischhauer}\ \emph {et~al.}(2005)\citenamefont
  {Fleischhauer}, \citenamefont {Imamoglu},\ and\ \citenamefont
  {Marangos}}]{Fleischhauer2005}%
  \BibitemOpen
  \bibfield  {author} {\bibinfo {author} {\bibfnamefont {M.}~\bibnamefont
  {Fleischhauer}}, \bibinfo {author} {\bibfnamefont {A.}~\bibnamefont
  {Imamoglu}},\ and\ \bibinfo {author} {\bibfnamefont {J.~P.}\ \bibnamefont
  {Marangos}},\ }\bibfield  {title} {\bibinfo {title} {Electromagnetically
  induced transparency: Optics in coherent media},\ }\href
  {https://doi.org/10.1103/RevModPhys.77.633} {\bibfield  {journal} {\bibinfo
  {journal} {Rev. Mod. Phys.}\ }\textbf {\bibinfo {volume} {77}},\ \bibinfo
  {pages} {633} (\bibinfo {year} {2005})}\BibitemShut {NoStop}%
\bibitem [{\citenamefont {Andersson}\ \emph {et~al.}(2020)\citenamefont
  {Andersson}, \citenamefont {Ekstr\"om},\ and\ \citenamefont
  {Delsing}}]{Andersson2020}%
  \BibitemOpen
  \bibfield  {author} {\bibinfo {author} {\bibfnamefont {G.}~\bibnamefont
  {Andersson}}, \bibinfo {author} {\bibfnamefont {M.~K.}\ \bibnamefont
  {Ekstr\"om}},\ and\ \bibinfo {author} {\bibfnamefont {P.}~\bibnamefont
  {Delsing}},\ }\bibfield  {title} {\bibinfo {title} {Electromagnetically
  induced acoustic transparency with a superconducting circuit},\ }\href
  {https://doi.org/10.1103/PhysRevLett.124.240402} {\bibfield  {journal}
  {\bibinfo  {journal} {Phys. Rev. Lett.}\ }\textbf {\bibinfo {volume} {124}},\
  \bibinfo {pages} {240402} (\bibinfo {year} {2020})}\BibitemShut {NoStop}%
\bibitem [{\citenamefont {Vadiraj}\ \emph {et~al.}(2021)\citenamefont
  {Vadiraj}, \citenamefont {Ask}, \citenamefont {McConkey}, \citenamefont
  {Nsanzineza}, \citenamefont {Chang}, \citenamefont {Kockum},\ and\
  \citenamefont {Wilson}}]{Vadiraj2021}%
  \BibitemOpen
  \bibfield  {author} {\bibinfo {author} {\bibfnamefont {A.~M.}\ \bibnamefont
  {Vadiraj}}, \bibinfo {author} {\bibfnamefont {A.}~\bibnamefont {Ask}},
  \bibinfo {author} {\bibfnamefont {T.~G.}\ \bibnamefont {McConkey}}, \bibinfo
  {author} {\bibfnamefont {I.}~\bibnamefont {Nsanzineza}}, \bibinfo {author}
  {\bibfnamefont {C.~W.~S.}\ \bibnamefont {Chang}}, \bibinfo {author}
  {\bibfnamefont {A.~F.}\ \bibnamefont {Kockum}},\ and\ \bibinfo {author}
  {\bibfnamefont {C.~M.}\ \bibnamefont {Wilson}},\ }\bibfield  {title}
  {\bibinfo {title} {Engineering the level structure of a giant artificial atom
  in waveguide quantum electrodynamics},\ }\href
  {https://doi.org/10.1103/PhysRevA.103.023710} {\bibfield  {journal} {\bibinfo
   {journal} {Phys. Rev. A}\ }\textbf {\bibinfo {volume} {103}},\ \bibinfo
  {pages} {023710} (\bibinfo {year} {2021})}\BibitemShut {NoStop}%
\bibitem [{\citenamefont {Shen}\ and\ \citenamefont {Fan}(2009)}]{Shen2009}%
  \BibitemOpen
  \bibfield  {author} {\bibinfo {author} {\bibfnamefont {J.-T.}\ \bibnamefont
  {Shen}}\ and\ \bibinfo {author} {\bibfnamefont {S.}~\bibnamefont {Fan}},\
  }\bibfield  {title} {\bibinfo {title} {Theory of single-photon transport in a
  single-mode waveguide. i. coupling to a cavity containing a two-level atom},\
  }\href {https://doi.org/10.1103/PhysRevA.79.023837} {\bibfield  {journal}
  {\bibinfo  {journal} {Phys. Rev. A}\ }\textbf {\bibinfo {volume} {79}},\
  \bibinfo {pages} {023837} (\bibinfo {year} {2009})}\BibitemShut {NoStop}%
\bibitem [{SM2()}]{SM2020}%
  \BibitemOpen
  \href@noop {} {\bibinfo {title} {Supplement materials}}\BibitemShut {NoStop}%
\bibitem [{\citenamefont {Jia}\ \emph {et~al.}(2017)\citenamefont {Jia},
  \citenamefont {Wang},\ and\ \citenamefont {Liu}}]{Jia2017}%
  \BibitemOpen
  \bibfield  {author} {\bibinfo {author} {\bibfnamefont {W.~Z.}\ \bibnamefont
  {Jia}}, \bibinfo {author} {\bibfnamefont {Y.~W.}\ \bibnamefont {Wang}},\ and\
  \bibinfo {author} {\bibfnamefont {Y.-x.}\ \bibnamefont {Liu}},\ }\bibfield
  {title} {\bibinfo {title} {Efficient single-photon frequency conversion in
  the microwave domain using superconducting quantum circuits},\ }\href
  {https://doi.org/10.1103/PhysRevA.96.053832} {\bibfield  {journal} {\bibinfo
  {journal} {Phys. Rev. A}\ }\textbf {\bibinfo {volume} {96}},\ \bibinfo
  {pages} {053832} (\bibinfo {year} {2017})}\BibitemShut {NoStop}%
\bibitem [{\citenamefont {Bradford}\ and\ \citenamefont
  {Shen}(2012)}]{Bradford2012}%
  \BibitemOpen
  \bibfield  {author} {\bibinfo {author} {\bibfnamefont {M.}~\bibnamefont
  {Bradford}}\ and\ \bibinfo {author} {\bibfnamefont {J.-T.}\ \bibnamefont
  {Shen}},\ }\bibfield  {title} {\bibinfo {title} {Single-photon frequency
  conversion by exploiting quantum interference},\ }\href
  {https://doi.org/10.1103/PhysRevA.85.043814} {\bibfield  {journal} {\bibinfo
  {journal} {Phys. Rev. A}\ }\textbf {\bibinfo {volume} {85}},\ \bibinfo
  {pages} {043814} (\bibinfo {year} {2012})}\BibitemShut {NoStop}%
\bibitem [{\citenamefont {{Gough}}\ and\ \citenamefont
  {{James}}(2009)}]{Gough2009}%
  \BibitemOpen
  \bibfield  {author} {\bibinfo {author} {\bibfnamefont {J.}~\bibnamefont
  {{Gough}}}\ and\ \bibinfo {author} {\bibfnamefont {M.~R.}\ \bibnamefont
  {{James}}},\ }\bibfield  {title} {\bibinfo {title} {The series product and
  its application to quantum feedforward and feedback networks},\ }\href
  {https://doi.org/10.1109/TAC.2009.2031205} {\bibfield  {journal} {\bibinfo
  {journal} {IEEE Trans. Autom. Control}\ }\textbf {\bibinfo {volume} {54}},\
  \bibinfo {pages} {2530} (\bibinfo {year} {2009})}\BibitemShut {NoStop}%
\bibitem [{\citenamefont {Manucharyan}\ \emph {et~al.}(2009)\citenamefont
  {Manucharyan}, \citenamefont {Koch}, \citenamefont {Glazman},\ and\
  \citenamefont {Devoret}}]{Manucharyan2013}%
  \BibitemOpen
  \bibfield  {author} {\bibinfo {author} {\bibfnamefont {V.~E.}\ \bibnamefont
  {Manucharyan}}, \bibinfo {author} {\bibfnamefont {J.}~\bibnamefont {Koch}},
  \bibinfo {author} {\bibfnamefont {L.~I.}\ \bibnamefont {Glazman}},\ and\
  \bibinfo {author} {\bibfnamefont {M.~H.}\ \bibnamefont {Devoret}},\
  }\bibfield  {title} {\bibinfo {title} {Fluxonium: Single cooper-pair circuit
  free of charge offsets},\ }\href {https://doi.org/10.1126/science.1175552}
  {\bibfield  {journal} {\bibinfo  {journal} {Science}\ }\textbf {\bibinfo
  {volume} {326}},\ \bibinfo {pages} {113} (\bibinfo {year}
  {2009})}\BibitemShut {NoStop}%
\bibitem [{\citenamefont {Abdumalikov}\ \emph {et~al.}(2010)\citenamefont
  {Abdumalikov}, \citenamefont {Astafiev}, \citenamefont {Zagoskin},
  \citenamefont {Pashkin}, \citenamefont {Nakamura},\ and\ \citenamefont
  {Tsai}}]{Abdumalikov2010}%
  \BibitemOpen
  \bibfield  {author} {\bibinfo {author} {\bibfnamefont {A.~A.}\ \bibnamefont
  {Abdumalikov}}, \bibinfo {author} {\bibfnamefont {O.}~\bibnamefont
  {Astafiev}}, \bibinfo {author} {\bibfnamefont {A.~M.}\ \bibnamefont
  {Zagoskin}}, \bibinfo {author} {\bibfnamefont {Y.~A.}\ \bibnamefont
  {Pashkin}}, \bibinfo {author} {\bibfnamefont {Y.}~\bibnamefont {Nakamura}},\
  and\ \bibinfo {author} {\bibfnamefont {J.~S.}\ \bibnamefont {Tsai}},\
  }\bibfield  {title} {\bibinfo {title} {{Electromagnetically induced
  transparency on a single artificial atom}},\ }\href
  {https://doi.org/10.1103/physrevlett.104.193601} {\bibfield  {journal}
  {\bibinfo  {journal} {Phys. Rev. Lett.}\ }\textbf {\bibinfo {volume} {104}},\
  \bibinfo {pages} {193601} (\bibinfo {year} {2010})}\BibitemShut {NoStop}%
\bibitem [{\citenamefont {Novikov}\ \emph {et~al.}(2016)\citenamefont
  {Novikov}, \citenamefont {Sweeney}, \citenamefont {Robinson}, \citenamefont
  {Premaratne}, \citenamefont {Suri}, \citenamefont {Wellstood},\ and\
  \citenamefont {Palmer}}]{Novikov2016}%
  \BibitemOpen
  \bibfield  {author} {\bibinfo {author} {\bibfnamefont {S.}~\bibnamefont
  {Novikov}}, \bibinfo {author} {\bibfnamefont {T.}~\bibnamefont {Sweeney}},
  \bibinfo {author} {\bibfnamefont {J.~E.}\ \bibnamefont {Robinson}}, \bibinfo
  {author} {\bibfnamefont {S.~P.}\ \bibnamefont {Premaratne}}, \bibinfo
  {author} {\bibfnamefont {B.}~\bibnamefont {Suri}}, \bibinfo {author}
  {\bibfnamefont {F.~C.}\ \bibnamefont {Wellstood}},\ and\ \bibinfo {author}
  {\bibfnamefont {B.~S.}\ \bibnamefont {Palmer}},\ }\bibfield  {title}
  {\bibinfo {title} {{Raman coherence in a circuit quantum electrodynamics
  lambda system}},\ }\href {https://doi.org/10.1038/nphys3537} {\bibfield
  {journal} {\bibinfo  {journal} {Nat. Phys.}\ }\textbf {\bibinfo {volume}
  {12}},\ \bibinfo {pages} {75} (\bibinfo {year} {2016})}\BibitemShut {NoStop}%
\bibitem [{\citenamefont {Long}\ \emph {et~al.}(2018)\citenamefont {Long},
  \citenamefont {Ku}, \citenamefont {Wu}, \citenamefont {Gu}, \citenamefont
  {Lake}, \citenamefont {Bal}, \citenamefont {Liu},\ and\ \citenamefont
  {Pappas}}]{Long2018}%
  \BibitemOpen
  \bibfield  {author} {\bibinfo {author} {\bibfnamefont {J.}~\bibnamefont
  {Long}}, \bibinfo {author} {\bibfnamefont {H.~S.}\ \bibnamefont {Ku}},
  \bibinfo {author} {\bibfnamefont {X.}~\bibnamefont {Wu}}, \bibinfo {author}
  {\bibfnamefont {X.}~\bibnamefont {Gu}}, \bibinfo {author} {\bibfnamefont
  {R.~E.}\ \bibnamefont {Lake}}, \bibinfo {author} {\bibfnamefont
  {M.}~\bibnamefont {Bal}}, \bibinfo {author} {\bibfnamefont {Y.-x.}\
  \bibnamefont {Liu}},\ and\ \bibinfo {author} {\bibfnamefont {D.~P.}\
  \bibnamefont {Pappas}},\ }\bibfield  {title} {\bibinfo {title}
  {{Electromagnetically induced transparency in circuit quantum electrodynamics
  with nested polariton states}},\ }\href
  {https://doi.org/10.1103/physrevlett.120.083602} {\bibfield  {journal}
  {\bibinfo  {journal} {Phys. Rev. Lett.}\ }\textbf {\bibinfo {volume} {120}},\
  \bibinfo {pages} {083602} (\bibinfo {year} {2018})}\BibitemShut {NoStop}%
\bibitem [{\citenamefont {Zhu}\ and\ \citenamefont {Jia}(2019)}]{Zhu2019}%
  \BibitemOpen
  \bibfield  {author} {\bibinfo {author} {\bibfnamefont {Y.~T.}\ \bibnamefont
  {Zhu}}\ and\ \bibinfo {author} {\bibfnamefont {W.~Z.}\ \bibnamefont {Jia}},\
  }\bibfield  {title} {\bibinfo {title} {Single-photon quantum router in the
  microwave regime utilizing double superconducting resonators with tunable
  coupling},\ }\href {https://doi.org/10.1103/PhysRevA.99.063815} {\bibfield
  {journal} {\bibinfo  {journal} {Phys. Rev. A}\ }\textbf {\bibinfo {volume}
  {99}},\ \bibinfo {pages} {063815} (\bibinfo {year} {2019})}\BibitemShut
  {NoStop}%
\bibitem [{\citenamefont {Autler}\ and\ \citenamefont
  {Townes}(1955)}]{Autler1955}%
  \BibitemOpen
  \bibfield  {author} {\bibinfo {author} {\bibfnamefont {S.~H.}\ \bibnamefont
  {Autler}}\ and\ \bibinfo {author} {\bibfnamefont {C.~H.}\ \bibnamefont
  {Townes}},\ }\bibfield  {title} {\bibinfo {title} {Stark effect in rapidly
  varying fields},\ }\href {https://doi.org/10.1103/PhysRev.100.703} {\bibfield
   {journal} {\bibinfo  {journal} {Phys. Rev.}\ }\textbf {\bibinfo {volume}
  {100}},\ \bibinfo {pages} {703} (\bibinfo {year} {1955})}\BibitemShut
  {NoStop}%
\bibitem [{\citenamefont {Fleischhauer}\ \emph {et~al.}(2000)\citenamefont
  {Fleischhauer}, \citenamefont {Matsko},\ and\ \citenamefont
  {Scully}}]{Fleischhauer2000}%
  \BibitemOpen
  \bibfield  {author} {\bibinfo {author} {\bibfnamefont {M.}~\bibnamefont
  {Fleischhauer}}, \bibinfo {author} {\bibfnamefont {A.~B.}\ \bibnamefont
  {Matsko}},\ and\ \bibinfo {author} {\bibfnamefont {M.~O.}\ \bibnamefont
  {Scully}},\ }\bibfield  {title} {\bibinfo {title} {Quantum limit of optical
  magnetometry in the presence of ac stark shifts},\ }\href
  {https://doi.org/10.1103/PhysRevA.62.013808} {\bibfield  {journal} {\bibinfo
  {journal} {Phys. Rev. A}\ }\textbf {\bibinfo {volume} {62}},\ \bibinfo
  {pages} {013808} (\bibinfo {year} {2000})}\BibitemShut {NoStop}%
\bibitem [{\citenamefont {Meyer}\ \emph {et~al.}(2021)\citenamefont {Meyer},
  \citenamefont {O'Brien}, \citenamefont {Fahey}, \citenamefont {Cox},\ and\
  \citenamefont {Kunz}}]{Meyer2021}%
  \BibitemOpen
  \bibfield  {author} {\bibinfo {author} {\bibfnamefont {D.~H.}\ \bibnamefont
  {Meyer}}, \bibinfo {author} {\bibfnamefont {C.}~\bibnamefont {O'Brien}},
  \bibinfo {author} {\bibfnamefont {D.~P.}\ \bibnamefont {Fahey}}, \bibinfo
  {author} {\bibfnamefont {K.~C.}\ \bibnamefont {Cox}},\ and\ \bibinfo {author}
  {\bibfnamefont {P.~D.}\ \bibnamefont {Kunz}},\ }\bibfield  {title} {\bibinfo
  {title} {Optimal atomic quantum sensing using
  electromagnetically-induced-transparency readout},\ }\href
  {https://doi.org/10.1103/PhysRevA.104.043103} {\bibfield  {journal} {\bibinfo
   {journal} {Phys. Rev. A}\ }\textbf {\bibinfo {volume} {104}},\ \bibinfo
  {pages} {043103} (\bibinfo {year} {2021})}\BibitemShut {NoStop}%
\bibitem [{\citenamefont {Breuer}\ \emph {et~al.}(2009)\citenamefont {Breuer},
  \citenamefont {Laine},\ and\ \citenamefont {Piilo}}]{Breuer2009}%
  \BibitemOpen
  \bibfield  {author} {\bibinfo {author} {\bibfnamefont {H.-P.}\ \bibnamefont
  {Breuer}}, \bibinfo {author} {\bibfnamefont {E.-M.}\ \bibnamefont {Laine}},\
  and\ \bibinfo {author} {\bibfnamefont {J.}~\bibnamefont {Piilo}},\ }\bibfield
   {title} {\bibinfo {title} {Measure for the degree of non-{M}arkovian
  behavior of quantum processes in open systems},\ }\href
  {https://doi.org/10.1103/PhysRevLett.103.210401} {\bibfield  {journal}
  {\bibinfo  {journal} {Phys. Rev. Lett.}\ }\textbf {\bibinfo {volume} {103}},\
  \bibinfo {pages} {210401} (\bibinfo {year} {2009})}\BibitemShut {NoStop}%
\bibitem [{\citenamefont {Laine}\ \emph {et~al.}(2010)\citenamefont {Laine},
  \citenamefont {Piilo},\ and\ \citenamefont {Breuer}}]{Laine2010}%
  \BibitemOpen
  \bibfield  {author} {\bibinfo {author} {\bibfnamefont {E.-M.}\ \bibnamefont
  {Laine}}, \bibinfo {author} {\bibfnamefont {J.}~\bibnamefont {Piilo}},\ and\
  \bibinfo {author} {\bibfnamefont {H.-P.}\ \bibnamefont {Breuer}},\ }\bibfield
   {title} {\bibinfo {title} {Measure for the non-{M}arkovianity of quantum
  processes},\ }\href {https://doi.org/10.1103/PhysRevA.81.062115} {\bibfield
  {journal} {\bibinfo  {journal} {Phys. Rev. A}\ }\textbf {\bibinfo {volume}
  {81}},\ \bibinfo {pages} {062115} (\bibinfo {year} {2010})}\BibitemShut
  {NoStop}%
\bibitem [{\citenamefont {Hall}\ \emph {et~al.}(2014)\citenamefont {Hall},
  \citenamefont {Cresser}, \citenamefont {Li},\ and\ \citenamefont
  {Andersson}}]{Hall2014}%
  \BibitemOpen
  \bibfield  {author} {\bibinfo {author} {\bibfnamefont {M.~J.~W.}\
  \bibnamefont {Hall}}, \bibinfo {author} {\bibfnamefont {J.~D.}\ \bibnamefont
  {Cresser}}, \bibinfo {author} {\bibfnamefont {L.}~\bibnamefont {Li}},\ and\
  \bibinfo {author} {\bibfnamefont {E.}~\bibnamefont {Andersson}},\ }\bibfield
  {title} {\bibinfo {title} {Canonical form of master equations and
  characterization of non-{M}arkovianity},\ }\href
  {https://doi.org/10.1103/PhysRevA.89.042120} {\bibfield  {journal} {\bibinfo
  {journal} {Phys. Rev. A}\ }\textbf {\bibinfo {volume} {89}},\ \bibinfo
  {pages} {042120} (\bibinfo {year} {2014})}\BibitemShut {NoStop}%
\bibitem [{\citenamefont {Breuer}\ \emph {et~al.}(2016)\citenamefont {Breuer},
  \citenamefont {Laine}, \citenamefont {Piilo},\ and\ \citenamefont
  {Vacchini}}]{Breuer2016}%
  \BibitemOpen
  \bibfield  {author} {\bibinfo {author} {\bibfnamefont {H.-P.}\ \bibnamefont
  {Breuer}}, \bibinfo {author} {\bibfnamefont {E.-M.}\ \bibnamefont {Laine}},
  \bibinfo {author} {\bibfnamefont {J.}~\bibnamefont {Piilo}},\ and\ \bibinfo
  {author} {\bibfnamefont {B.}~\bibnamefont {Vacchini}},\ }\bibfield  {title}
  {\bibinfo {title} {{Colloquium: Non-Markovian dynamics in open quantum
  systems}},\ }\href {https://doi.org/10.1103/revmodphys.88.021002} {\bibfield
  {journal} {\bibinfo  {journal} {Rev. Mod. Phys.}\ }\textbf {\bibinfo {volume}
  {88}},\ \bibinfo {pages} {021002} (\bibinfo {year} {2016})}\BibitemShut
  {NoStop}%
\bibitem [{\citenamefont {Xue}\ \emph {et~al.}(2012)\citenamefont {Xue},
  \citenamefont {Wu}, \citenamefont {Zhang}, \citenamefont {Zhang},
  \citenamefont {Li},\ and\ \citenamefont {Tarn}}]{Xue2012}%
  \BibitemOpen
  \bibfield  {author} {\bibinfo {author} {\bibfnamefont {S.-B.}\ \bibnamefont
  {Xue}}, \bibinfo {author} {\bibfnamefont {R.-B.}\ \bibnamefont {Wu}},
  \bibinfo {author} {\bibfnamefont {W.-M.}\ \bibnamefont {Zhang}}, \bibinfo
  {author} {\bibfnamefont {J.}~\bibnamefont {Zhang}}, \bibinfo {author}
  {\bibfnamefont {C.-W.}\ \bibnamefont {Li}},\ and\ \bibinfo {author}
  {\bibfnamefont {T.-J.}\ \bibnamefont {Tarn}},\ }\bibfield  {title} {\bibinfo
  {title} {Decoherence suppression via non-{M}arkovian coherent feedback
  control},\ }\href {https://doi.org/10.1103/PhysRevA.86.052304} {\bibfield
  {journal} {\bibinfo  {journal} {Phys. Rev. A}\ }\textbf {\bibinfo {volume}
  {86}},\ \bibinfo {pages} {052304} (\bibinfo {year} {2012})}\BibitemShut
  {NoStop}%
\bibitem [{\citenamefont {Zhang}\ \emph {et~al.}(2012)\citenamefont {Zhang},
  \citenamefont {Lo}, \citenamefont {Xiong}, \citenamefont {Tu},\ and\
  \citenamefont {Nori}}]{Zhang2012}%
  \BibitemOpen
  \bibfield  {author} {\bibinfo {author} {\bibfnamefont {W.-M.}\ \bibnamefont
  {Zhang}}, \bibinfo {author} {\bibfnamefont {P.-Y.}\ \bibnamefont {Lo}},
  \bibinfo {author} {\bibfnamefont {H.-N.}\ \bibnamefont {Xiong}}, \bibinfo
  {author} {\bibfnamefont {M.~W.-Y.}\ \bibnamefont {Tu}},\ and\ \bibinfo
  {author} {\bibfnamefont {F.}~\bibnamefont {Nori}},\ }\bibfield  {title}
  {\bibinfo {title} {General non-{M}arkovian dynamics of open quantum
  systems},\ }\href {https://doi.org/10.1103/PhysRevLett.109.170402} {\bibfield
   {journal} {\bibinfo  {journal} {Phys. Rev. Lett.}\ }\textbf {\bibinfo
  {volume} {109}},\ \bibinfo {pages} {170402} (\bibinfo {year}
  {2012})}\BibitemShut {NoStop}%
\bibitem [{\citenamefont {Fang}\ and\ \citenamefont
  {Baranger}(2015)}]{Fang2015}%
  \BibitemOpen
  \bibfield  {author} {\bibinfo {author} {\bibfnamefont {Y.-L.~L.}\
  \bibnamefont {Fang}}\ and\ \bibinfo {author} {\bibfnamefont {H.~U.}\
  \bibnamefont {Baranger}},\ }\bibfield  {title} {\bibinfo {title} {Waveguide
  qed: Power spectra and correlations of two photons scattered off multiple
  distant qubits and a mirror},\ }\href
  {https://doi.org/10.1103/PhysRevA.91.053845} {\bibfield  {journal} {\bibinfo
  {journal} {Phys. Rev. A}\ }\textbf {\bibinfo {volume} {91}},\ \bibinfo
  {pages} {053845} (\bibinfo {year} {2015})}\BibitemShut {NoStop}%
\bibitem [{\citenamefont {Calaj\'o}\ \emph {et~al.}(2019)\citenamefont
  {Calaj\'o}, \citenamefont {Fang}, \citenamefont {Baranger},\ and\
  \citenamefont {Ciccarello}}]{Calajo2019}%
  \BibitemOpen
  \bibfield  {author} {\bibinfo {author} {\bibfnamefont {G.}~\bibnamefont
  {Calaj\'o}}, \bibinfo {author} {\bibfnamefont {Y.-L.~L.}\ \bibnamefont
  {Fang}}, \bibinfo {author} {\bibfnamefont {H.~U.}\ \bibnamefont {Baranger}},\
  and\ \bibinfo {author} {\bibfnamefont {F.}~\bibnamefont {Ciccarello}},\
  }\bibfield  {title} {\bibinfo {title} {Exciting a bound state in the
  continuum through multiphoton scattering plus delayed quantum feedback},\
  }\href {https://doi.org/10.1103/PhysRevLett.122.073601} {\bibfield  {journal}
  {\bibinfo  {journal} {Phys. Rev. Lett.}\ }\textbf {\bibinfo {volume} {122}},\
  \bibinfo {pages} {073601} (\bibinfo {year} {2019})}\BibitemShut {NoStop}%
\bibitem [{\citenamefont {Delsing}\ \emph {et~al.}(2019)\citenamefont
  {Delsing}, \citenamefont {Cleland}, \citenamefont {Schuetz}, \citenamefont
  {KnÃ¶rzer}, \citenamefont {Giedke}, \citenamefont {Cirac}, \citenamefont
  {Srinivasan}, \citenamefont {Wu}, \citenamefont {Balram}, \citenamefont
  {BÃ¤uerle}, \citenamefont {Meunier}, \citenamefont {Ford}, \citenamefont
  {Santos}, \citenamefont {Cerda-M{\'{e}}ndez}, \citenamefont {Wang},
  \citenamefont {Krenner}, \citenamefont {Nysten}, \citenamefont {Wei{\ss}},
  \citenamefont {Nash}, \citenamefont {Thevenard}, \citenamefont {Gourdon},
  \citenamefont {Rovillain}, \citenamefont {Marangolo}, \citenamefont
  {Duquesne}, \citenamefont {Fischerauer}, \citenamefont {Ruile}, \citenamefont
  {Reiner}, \citenamefont {Paschke}, \citenamefont {Denysenko}, \citenamefont
  {Volkmer}, \citenamefont {Wixforth}, \citenamefont {Bruus}, \citenamefont
  {Wiklund}, \citenamefont {Reboud}, \citenamefont {Cooper}, \citenamefont
  {Fu}, \citenamefont {Brugger}, \citenamefont {Rehfeldt},\ and\ \citenamefont
  {Westerhausen}}]{Delsing2019}%
  \BibitemOpen
  \bibfield  {author} {\bibinfo {author} {\bibfnamefont {P.}~\bibnamefont
  {Delsing}}, \bibinfo {author} {\bibfnamefont {A.~N.}\ \bibnamefont
  {Cleland}}, \bibinfo {author} {\bibfnamefont {M.~J.~A.}\ \bibnamefont
  {Schuetz}}, \bibinfo {author} {\bibfnamefont {J.}~\bibnamefont {KnÃ¶rzer}},
  \bibinfo {author} {\bibfnamefont {G.}~\bibnamefont {Giedke}}, \bibinfo
  {author} {\bibfnamefont {J.~I.}\ \bibnamefont {Cirac}}, \bibinfo {author}
  {\bibfnamefont {K.}~\bibnamefont {Srinivasan}}, \bibinfo {author}
  {\bibfnamefont {M.}~\bibnamefont {Wu}}, \bibinfo {author} {\bibfnamefont
  {K.~C.}\ \bibnamefont {Balram}}, \bibinfo {author} {\bibfnamefont
  {C.}~\bibnamefont {BÃ¤uerle}}, \bibinfo {author} {\bibfnamefont
  {T.}~\bibnamefont {Meunier}}, \bibinfo {author} {\bibfnamefont {C.~J.~B.}\
  \bibnamefont {Ford}}, \bibinfo {author} {\bibfnamefont {P.~V.}\ \bibnamefont
  {Santos}}, \bibinfo {author} {\bibfnamefont {E.}~\bibnamefont
  {Cerda-M{\'{e}}ndez}}, \bibinfo {author} {\bibfnamefont {H.}~\bibnamefont
  {Wang}}, \bibinfo {author} {\bibfnamefont {H.~J.}\ \bibnamefont {Krenner}},
  \bibinfo {author} {\bibfnamefont {E.~D.~S.}\ \bibnamefont {Nysten}}, \bibinfo
  {author} {\bibfnamefont {M.}~\bibnamefont {Wei{\ss}}}, \bibinfo {author}
  {\bibfnamefont {G.~R.}\ \bibnamefont {Nash}}, \bibinfo {author}
  {\bibfnamefont {L.}~\bibnamefont {Thevenard}}, \bibinfo {author}
  {\bibfnamefont {C.}~\bibnamefont {Gourdon}}, \bibinfo {author} {\bibfnamefont
  {P.}~\bibnamefont {Rovillain}}, \bibinfo {author} {\bibfnamefont
  {M.}~\bibnamefont {Marangolo}}, \bibinfo {author} {\bibfnamefont {J.-Y.}\
  \bibnamefont {Duquesne}}, \bibinfo {author} {\bibfnamefont {G.}~\bibnamefont
  {Fischerauer}}, \bibinfo {author} {\bibfnamefont {W.}~\bibnamefont {Ruile}},
  \bibinfo {author} {\bibfnamefont {A.}~\bibnamefont {Reiner}}, \bibinfo
  {author} {\bibfnamefont {B.}~\bibnamefont {Paschke}}, \bibinfo {author}
  {\bibfnamefont {D.}~\bibnamefont {Denysenko}}, \bibinfo {author}
  {\bibfnamefont {D.}~\bibnamefont {Volkmer}}, \bibinfo {author} {\bibfnamefont
  {A.}~\bibnamefont {Wixforth}}, \bibinfo {author} {\bibfnamefont
  {H.}~\bibnamefont {Bruus}}, \bibinfo {author} {\bibfnamefont
  {M.}~\bibnamefont {Wiklund}}, \bibinfo {author} {\bibfnamefont
  {J.}~\bibnamefont {Reboud}}, \bibinfo {author} {\bibfnamefont {J.~M.}\
  \bibnamefont {Cooper}}, \bibinfo {author} {\bibfnamefont {Y.}~\bibnamefont
  {Fu}}, \bibinfo {author} {\bibfnamefont {M.~S.}\ \bibnamefont {Brugger}},
  \bibinfo {author} {\bibfnamefont {F.}~\bibnamefont {Rehfeldt}},\ and\
  \bibinfo {author} {\bibfnamefont {C.}~\bibnamefont {Westerhausen}},\
  }\bibfield  {title} {\bibinfo {title} {The 2019 surface acoustic waves
  roadmap},\ }\href {https://doi.org/10.1088/1361-6463/ab1b04} {\bibfield
  {journal} {\bibinfo  {journal} {Journal of Physics D: Applied Physics}\
  }\textbf {\bibinfo {volume} {52}},\ \bibinfo {pages} {353001} (\bibinfo
  {year} {2019})}\BibitemShut {NoStop}%
\bibitem [{\citenamefont {Xue}\ \emph {et~al.}(2017{\natexlab{a}})\citenamefont
  {Xue}, \citenamefont {Wu}, \citenamefont {Hush},\ and\ \citenamefont
  {Tarn}}]{Xue2017A}%
  \BibitemOpen
  \bibfield  {author} {\bibinfo {author} {\bibfnamefont {S.}~\bibnamefont
  {Xue}}, \bibinfo {author} {\bibfnamefont {R.}~\bibnamefont {Wu}}, \bibinfo
  {author} {\bibfnamefont {M.~R.}\ \bibnamefont {Hush}},\ and\ \bibinfo
  {author} {\bibfnamefont {T.-J.}\ \bibnamefont {Tarn}},\ }\bibfield  {title}
  {\bibinfo {title} {{Non-{M}arkovian coherent feedback control of quantum dot
  systems}},\ }\href {https://doi.org/10.1088/2058-9565/aa6125} {\bibfield
  {journal} {\bibinfo  {journal} {Quantum Sci. Technol.}\ }\textbf {\bibinfo
  {volume} {2}},\ \bibinfo {pages} {014002} (\bibinfo {year}
  {2017}{\natexlab{a}})}\BibitemShut {NoStop}%
\bibitem [{\citenamefont {Xue}\ \emph {et~al.}(2017{\natexlab{b}})\citenamefont
  {Xue}, \citenamefont {Hush},\ and\ \citenamefont {Petersen}}]{Xue2017B}%
  \BibitemOpen
  \bibfield  {author} {\bibinfo {author} {\bibfnamefont {S.}~\bibnamefont
  {Xue}}, \bibinfo {author} {\bibfnamefont {M.~R.}\ \bibnamefont {Hush}},\ and\
  \bibinfo {author} {\bibfnamefont {I.~R.}\ \bibnamefont {Petersen}},\
  }\bibfield  {title} {\bibinfo {title} {Feedback tracking control of
  non-{M}arkovian quantum systems},\ }\href
  {https://doi.org/10.1109/TCST.2016.2614834} {\bibfield  {journal} {\bibinfo
  {journal} {IEEE Trans. Control Syst. Technol.}\ }\textbf {\bibinfo {volume}
  {25}},\ \bibinfo {pages} {1552} (\bibinfo {year}
  {2017}{\natexlab{b}})}\BibitemShut {NoStop}%
\bibitem [{\citenamefont {Xue}\ \emph {et~al.}(2020)\citenamefont {Xue},
  \citenamefont {Nguyen}, \citenamefont {James}, \citenamefont {Shabani},
  \citenamefont {Ugrinovskii},\ and\ \citenamefont {Petersen}}]{Xue2020}%
  \BibitemOpen
  \bibfield  {author} {\bibinfo {author} {\bibfnamefont {S.}~\bibnamefont
  {Xue}}, \bibinfo {author} {\bibfnamefont {T.}~\bibnamefont {Nguyen}},
  \bibinfo {author} {\bibfnamefont {M.~R.}\ \bibnamefont {James}}, \bibinfo
  {author} {\bibfnamefont {A.}~\bibnamefont {Shabani}}, \bibinfo {author}
  {\bibfnamefont {V.}~\bibnamefont {Ugrinovskii}},\ and\ \bibinfo {author}
  {\bibfnamefont {I.~R.}\ \bibnamefont {Petersen}},\ }\bibfield  {title}
  {\bibinfo {title} {Modeling for non-markovian quantum systems},\ }\href
  {https://doi.org/10.1109/TCST.2019.2935421} {\bibfield  {journal} {\bibinfo
  {journal} {IEEE Trans. Control Syst. Technol.}\ }\textbf {\bibinfo {volume}
  {28}},\ \bibinfo {pages} {2564} (\bibinfo {year} {2020})}\BibitemShut
  {NoStop}%
\bibitem [{\citenamefont {Hoi}\ \emph {et~al.}(2011)\citenamefont {Hoi},
  \citenamefont {Wilson}, \citenamefont {Johansson}, \citenamefont {Palomaki},
  \citenamefont {Peropadre},\ and\ \citenamefont {Delsing}}]{Hoi2011}%
  \BibitemOpen
  \bibfield  {author} {\bibinfo {author} {\bibfnamefont {I.-C.}\ \bibnamefont
  {Hoi}}, \bibinfo {author} {\bibfnamefont {C.~M.}\ \bibnamefont {Wilson}},
  \bibinfo {author} {\bibfnamefont {G.}~\bibnamefont {Johansson}}, \bibinfo
  {author} {\bibfnamefont {T.}~\bibnamefont {Palomaki}}, \bibinfo {author}
  {\bibfnamefont {B.}~\bibnamefont {Peropadre}},\ and\ \bibinfo {author}
  {\bibfnamefont {P.}~\bibnamefont {Delsing}},\ }\bibfield  {title} {\bibinfo
  {title} {Demonstration of a single-photon router in the microwave regime},\
  }\href {https://doi.org/10.1103/PhysRevLett.107.073601} {\bibfield  {journal}
  {\bibinfo  {journal} {Phys. Rev. Lett.}\ }\textbf {\bibinfo {volume} {107}},\
  \bibinfo {pages} {073601} (\bibinfo {year} {2011})}\BibitemShut {NoStop}%
\bibitem [{\citenamefont {Afzelius}\ \emph {et~al.}(2015)\citenamefont
  {Afzelius}, \citenamefont {Gisin},\ and\ \citenamefont
  {Riedmatten}}]{Afzelius2015}%
  \BibitemOpen
  \bibfield  {author} {\bibinfo {author} {\bibfnamefont {M.}~\bibnamefont
  {Afzelius}}, \bibinfo {author} {\bibfnamefont {N.}~\bibnamefont {Gisin}},\
  and\ \bibinfo {author} {\bibfnamefont {H.}~\bibnamefont {Riedmatten}},\
  }\bibfield  {title} {\bibinfo {title} {{Quantum memory for photons}},\ }\href
  {https://doi.org/10.1063/pt.3.3021} {\bibfield  {journal} {\bibinfo
  {journal} {Phys. Today}\ }\textbf {\bibinfo {volume} {68}},\ \bibinfo {pages}
  {42} (\bibinfo {year} {2015})}\BibitemShut {NoStop}%
\bibitem [{\citenamefont {Kasapi}\ \emph {et~al.}(1995)\citenamefont {Kasapi},
  \citenamefont {Jain}, \citenamefont {Yin},\ and\ \citenamefont
  {Harris}}]{Kasapi1995}%
  \BibitemOpen
  \bibfield  {author} {\bibinfo {author} {\bibfnamefont {A.}~\bibnamefont
  {Kasapi}}, \bibinfo {author} {\bibfnamefont {M.}~\bibnamefont {Jain}},
  \bibinfo {author} {\bibfnamefont {G.~Y.}\ \bibnamefont {Yin}},\ and\ \bibinfo
  {author} {\bibfnamefont {S.~E.}\ \bibnamefont {Harris}},\ }\bibfield  {title}
  {\bibinfo {title} {Electromagnetically induced transparency: Propagation
  dynamics},\ }\href {https://doi.org/10.1103/PhysRevLett.74.2447} {\bibfield
  {journal} {\bibinfo  {journal} {Phys. Rev. Lett.}\ }\textbf {\bibinfo
  {volume} {74}},\ \bibinfo {pages} {2447} (\bibinfo {year}
  {1995})}\BibitemShut {NoStop}%
\bibitem [{\citenamefont {Xiao}\ \emph {et~al.}(1995)\citenamefont {Xiao},
  \citenamefont {Li}, \citenamefont {Jin},\ and\ \citenamefont
  {Gea-Banacloche}}]{Xiao1995}%
  \BibitemOpen
  \bibfield  {author} {\bibinfo {author} {\bibfnamefont {M.}~\bibnamefont
  {Xiao}}, \bibinfo {author} {\bibfnamefont {Y.-q.}\ \bibnamefont {Li}},
  \bibinfo {author} {\bibfnamefont {S.-z.}\ \bibnamefont {Jin}},\ and\ \bibinfo
  {author} {\bibfnamefont {J.}~\bibnamefont {Gea-Banacloche}},\ }\bibfield
  {title} {\bibinfo {title} {Measurement of dispersive properties of
  electromagnetically induced transparency in rubidium atoms},\ }\href
  {https://doi.org/10.1103/PhysRevLett.74.666} {\bibfield  {journal} {\bibinfo
  {journal} {Phys. Rev. Lett.}\ }\textbf {\bibinfo {volume} {74}},\ \bibinfo
  {pages} {666} (\bibinfo {year} {1995})}\BibitemShut {NoStop}%
\bibitem [{\citenamefont {Schmidt}\ and\ \citenamefont
  {Imamoglu}(1996)}]{Schmidt1996}%
  \BibitemOpen
  \bibfield  {author} {\bibinfo {author} {\bibfnamefont {H.}~\bibnamefont
  {Schmidt}}\ and\ \bibinfo {author} {\bibfnamefont {A.}~\bibnamefont
  {Imamoglu}},\ }\bibfield  {title} {\bibinfo {title} {Giant kerr
  nonlinearities obtained by electromagnetically induced transparency},\ }\href
  {https://doi.org/10.1364/OL.21.001936} {\bibfield  {journal} {\bibinfo
  {journal} {Opt. Lett.}\ }\textbf {\bibinfo {volume} {21}},\ \bibinfo {pages}
  {1936} (\bibinfo {year} {1996})}\BibitemShut {NoStop}%
\bibitem [{\citenamefont {Scully}\ \emph {et~al.}(1989)\citenamefont {Scully},
  \citenamefont {Zhu},\ and\ \citenamefont {Gavrielides}}]{Scully1989}%
  \BibitemOpen
  \bibfield  {author} {\bibinfo {author} {\bibfnamefont {M.~O.}\ \bibnamefont
  {Scully}}, \bibinfo {author} {\bibfnamefont {S.-Y.}\ \bibnamefont {Zhu}},\
  and\ \bibinfo {author} {\bibfnamefont {A.}~\bibnamefont {Gavrielides}},\
  }\bibfield  {title} {\bibinfo {title} {Degenerate quantum-beat laser: Lasing
  without inversion and inversion without lasing},\ }\href
  {https://doi.org/10.1103/PhysRevLett.62.2813} {\bibfield  {journal} {\bibinfo
   {journal} {Phys. Rev. Lett.}\ }\textbf {\bibinfo {volume} {62}},\ \bibinfo
  {pages} {2813} (\bibinfo {year} {1989})}\BibitemShut {NoStop}%
\bibitem [{\citenamefont {Fry}\ \emph {et~al.}(1993)\citenamefont {Fry},
  \citenamefont {Li}, \citenamefont {Nikonov}, \citenamefont {Padmabandu},
  \citenamefont {Scully}, \citenamefont {Smith}, \citenamefont {Tittel},
  \citenamefont {Wang}, \citenamefont {Wilkinson},\ and\ \citenamefont
  {Zhu}}]{Fry1993}%
  \BibitemOpen
  \bibfield  {author} {\bibinfo {author} {\bibfnamefont {E.~S.}\ \bibnamefont
  {Fry}}, \bibinfo {author} {\bibfnamefont {X.}~\bibnamefont {Li}}, \bibinfo
  {author} {\bibfnamefont {D.}~\bibnamefont {Nikonov}}, \bibinfo {author}
  {\bibfnamefont {G.~G.}\ \bibnamefont {Padmabandu}}, \bibinfo {author}
  {\bibfnamefont {M.~O.}\ \bibnamefont {Scully}}, \bibinfo {author}
  {\bibfnamefont {A.~V.}\ \bibnamefont {Smith}}, \bibinfo {author}
  {\bibfnamefont {F.~K.}\ \bibnamefont {Tittel}}, \bibinfo {author}
  {\bibfnamefont {C.}~\bibnamefont {Wang}}, \bibinfo {author} {\bibfnamefont
  {S.~R.}\ \bibnamefont {Wilkinson}},\ and\ \bibinfo {author} {\bibfnamefont
  {S.-Y.}\ \bibnamefont {Zhu}},\ }\bibfield  {title} {\bibinfo {title} {Atomic
  coherence effects within the sodium ${\mathit{d}}_{1}$ line: Lasing without
  inversion via population trapping},\ }\href
  {https://doi.org/10.1103/PhysRevLett.70.3235} {\bibfield  {journal} {\bibinfo
   {journal} {Phys. Rev. Lett.}\ }\textbf {\bibinfo {volume} {70}},\ \bibinfo
  {pages} {3235} (\bibinfo {year} {1993})}\BibitemShut {NoStop}%
\bibitem [{\citenamefont {Nottelmann}\ \emph {et~al.}(1993)\citenamefont
  {Nottelmann}, \citenamefont {Peters},\ and\ \citenamefont
  {Lange}}]{Nottelmann1993}%
  \BibitemOpen
  \bibfield  {author} {\bibinfo {author} {\bibfnamefont {A.}~\bibnamefont
  {Nottelmann}}, \bibinfo {author} {\bibfnamefont {C.}~\bibnamefont {Peters}},\
  and\ \bibinfo {author} {\bibfnamefont {W.}~\bibnamefont {Lange}},\ }\bibfield
   {title} {\bibinfo {title} {Inversionless amplification of picosecond pulses
  due to zeeman coherence},\ }\href
  {https://doi.org/10.1103/PhysRevLett.70.1783} {\bibfield  {journal} {\bibinfo
   {journal} {Phys. Rev. Lett.}\ }\textbf {\bibinfo {volume} {70}},\ \bibinfo
  {pages} {1783} (\bibinfo {year} {1993})}\BibitemShut {NoStop}%
\end{thebibliography}%

\end{document}